\documentclass[aps,pre,twocolumn]{revtex4}
\usepackage{bm}
\usepackage{graphicx}
\usepackage{amsmath}
\usepackage{amssymb}
\usepackage{siunitx}

\renewcommand{\H}{{\cal{H}}}
\renewcommand{\P}{{\cal{P}}}
\newcommand{\tu}{\tilde{u}}
\newcommand{\fr}[2]{{\textstyle\frac{#1}{#2}}}

\begin{document}

\title{Almost-dispersionless pulse transport in long quasiuniform
       spring-mass chains:  \\A new kind of Newton's cradle}

\author{Ruggero Vaia}
\email{ruggero.vaia@isc.cnr.it}
\affiliation{Istituto dei Sistemi Complessi, Consiglio Nazionale delle Ricerche,
             I-50019 Sesto Fiorentino, Italy}
\affiliation{Istituto Nazionale di Fisica Nucleare, Sezione di Firenze,
             I-50019 Sesto Fiorentino, Italy}

\date{\today}

\begin{abstract}
Almost-dispersionless pulse transfer between the extremal masses of a uniform harmonic spring-mass chain of arbitrary length can be induced by suitably modifying two masses and their spring's elastic constant at both extrema of the chain. It is shown that a deviation (or a pulse) imposed to the first mass gives rise to a wave packet that, after a time of the order of the chain length, almost perfectly reproduces the same deviation (pulse) at the opposite end, with an amplitude loss that is as small as 1.3\,\% in the infinite-length limit; such a dynamics can continue back and forth again for several times before dispersion cleared the effect. The underlying coherence mechanism is that the initial condition excites a bunch of normal modes with almost equal frequency spacing. This constitutes a possible mechanism for efficient energy transfer, e.g., in nanofabricated structures.
\end{abstract}



\maketitle

\section{Introduction}
\label{s.intro}

Newton's cradle is a toy that seems to have a perfect behavior, explained in terms of the conservation of energy and momentum in the pairwise elastic collision of equal spheres. However, such a na\"{\i}ve theory only holds for two spheres: For many spheres the underlying physics is complicated and needs a detailed analysis of the collision mechanism. For instance, observation gives the impression that the internal spheres do not move, so one could imagine that gluing them together would not affect the cradle. This is in contrast with the conservation laws, for only a fraction of the momentum would be transferred from the first colliding sphere; in any case, why Newton's cradle works, as it indeed does, is a well settled issue~\cite{HermannS1982,HutzlerDWM2004,Glendinning2011}.

\begin{figure}
\includegraphics[width=0.47\textwidth]{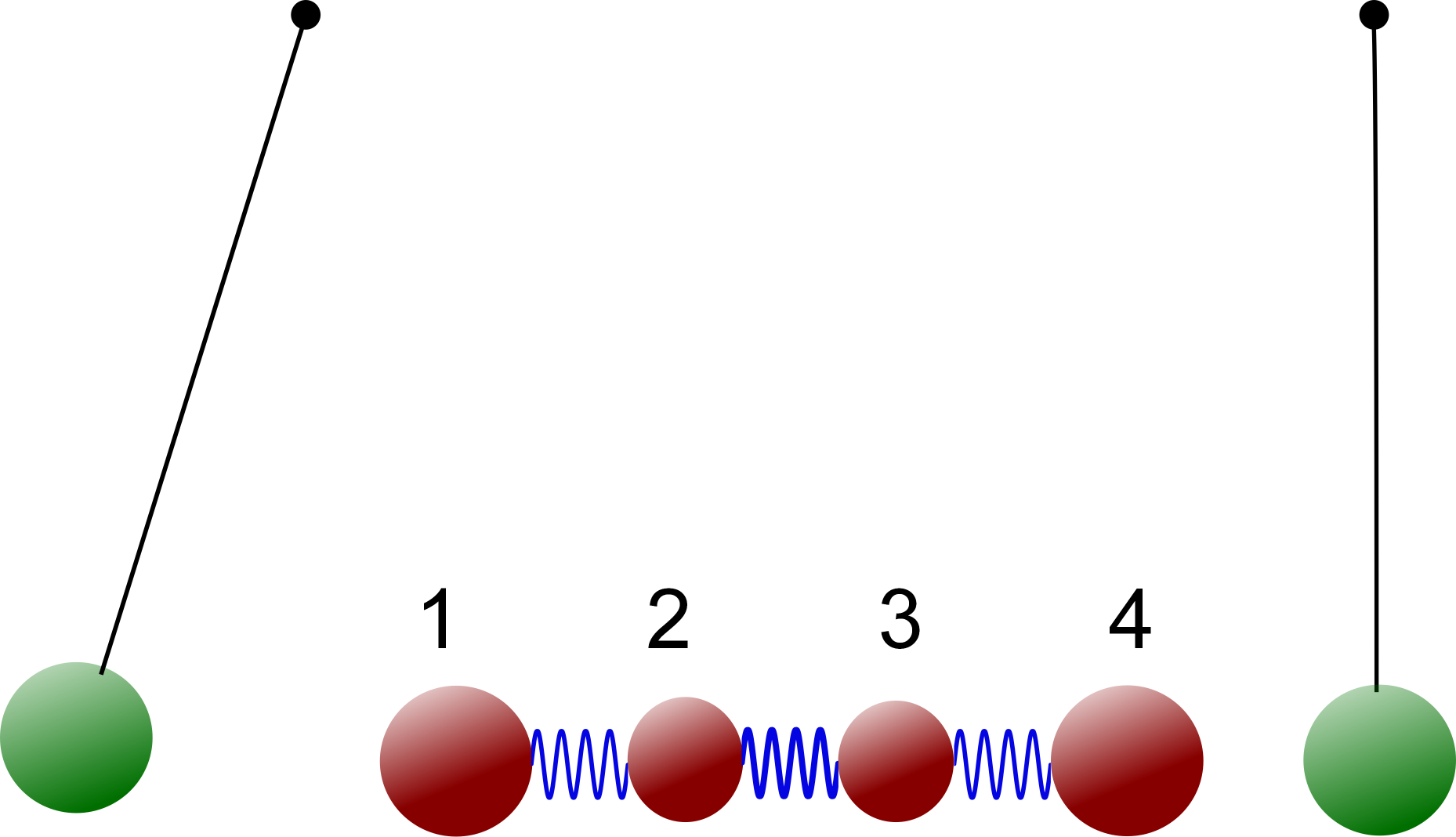}
\caption{A perfect cradle with a 4-mass chain. The first and fourth masses are 5/3 of the internal ones, while the external springs' constants are 5/6 of the internal one: This setup yields perfect end-to-end transmission (see Sec.~\ref{ss.N3N4}). The auxiliary (green) external masses, equal to the first and last ones, behave just like in a Newton's cradle, except for the finite transmission time. In Sec.~\ref{s.m1m2K12} it is shown that chains of arbitrary length can behave almost in the same way if the extremal mass and spring values are chosen according to Table~\ref{t.opt_rw}.}
	\label{f.cradle4}
\end{figure}

In this paper a different system is considered, namely a chain of $N$ masses connected by elastic massless springs, such as that shown in Fig.~\ref{f.cradle4}, looking for the possibility that it behaved in an analogous way, by this meaning that an initial ``pulse'' located on the first mass travels along the chain and reaches the last mass yielding a mirror image of the initial configuration. This requirement is far from trivial. The simplest choice, namely a uniform chain with identical masses and identical springs, must be ruled out, since it is easily proven that it cannot coherently transfer a pulse between its ends~\cite{RosasL2004}. Indeed, only if the frequencies of the normal modes involved in the dynamics had a greatest common divisor, i.e., they were integer multiples of a finite frequency $\delta\omega\equiv\pi/T$, then after a time $2T$ the system would be exactly back to the initial configuration and (provided the chain be mirror symmetric) at $T$ the initial pulse would be found at the opposite end. The uniform chain does not possess the above requisite, because its frequencies  $\omega_n\propto\sin(cn)$, with $c$ constant and $n$ integer, cannot have a common divisor: An initial pulse would undergo dispersion giving rise to a seemingly chaotic dynamics.

In order to achieve dispersionless pulse transmission, most studies have been considering nonlinear chains: A uniform array with suitable anharmonic terms can indeed support the propagation of localized wave packets, sometimes dubbed {\em breathers\,}~\cite{FlachW1998,RosasL2004,SarmientoRRL1999}. However, the goal of coherent transmission can be attained without invoking nonlinearity but rather allowing for a slight modification from uniformity. An optimal choice of the two masses and of the spring between them at both chain ends results in an almost-coherent dynamics, with the initial pulse traveling along the uniform bulk of the chain and reconstructing itself with high fidelity at the opposite end, thereafter bouncing back and forth many times, just as the Newton cradle does, with the difference that propagation along the chain takes a finite time, proportional to its length. The optimal masses and spring values depend on $N$, but, remarkably, the transmission quality is almost independent of $N$: Beyond $N\sim{60}$ it even improves for increasing $N$. For $N\to\infty$ the optimized parameters can be studied analytically, showing that the amplitude loss on transmission tends to 1.28\%, a result confirmed by a numerical approach, necessary for finite $N$.
Among many experimental papers, Ref.~\onlinecite{RammPH2014} considers a setup, involving an ion chain which is kicked at one end, that is quite similar to the model studied here. In this experiment it appears that the chain is well isolated from the environment, so that friction can be assumed to be negligible.

The classical spring-mass chain can be realized in alternative ways, such as a sequence of bars or disks connected by torsion wires, or an electrical circuit with capacitors and inductances, where the role of the masses is played by the moments of inertia or by the inductances, respectively. 

In Sec.~\ref{s.model} the spring-mass chain model is presented and described in terms of dimensionless parameters; the {\em transmission amplitude} is introduced as the relevant quality factor, which can be maximized to 100\,\% in the analytically solvable cases with $N\le5$. A general analytical approach to the normal modes of the quasiuniform chain for any $N$ is used in Sec.~\ref{s.algebra} to evaluate the main ingredient for calculating the transmission amplitude, namely the {\em mode density} describing how much the different normal modes are excited by the initial pulse. In Sec.~\ref{s.m1} the behavior of the optimized parameters is studied when varying one extremal mass only, while in Sec.~\ref{s.m1m2K12} the transmission is shown to be strongly improved by optimizing two extremal masses and their connecting spring. Sec.~\ref{s.concl} contains a summary of the achievements and suggestions for possible applications.

\section{Transmission in the spring-mass chain}
\label{s.model}

\subsection{The model}

Consider a chain of masses connected by springs whose Hamiltonian reads
\begin{equation}
\H = \sum_{i=1}^N \frac{P_i^2}{2m_i}
+ \frac12\sum_{i=0}^{N}K_{i,i{+}1}(Q_i-Q_{i{+}1})^2 ~,
\label{e.Horig}
\end{equation}
with $Q_0=Q_{N{+}1}=0$, i.e., the extremal springs $K_{01}$ and $K_{N,N{+}1}$ connect each one of the masses $m_1$ and $m_N$ to a fixed point, say a ``wall''. Performing the canonical transformation to mass-weighted variables, $P_i=p_i\sqrt{m_i}$ and $Q_i=q_i/\sqrt{m_i}$, the Hamiltonian turns into
\begin{equation}
\H =  \frac12\sum_{i=1}^N p_i^2 + \frac{K}{2m}\sum_{i,j=1}^N B_{ij}\,q_iq_j~,
\label{e.H}
\end{equation}
where $m$ and $K$ are the ``typical'' mass and elastic constant, and the dimensionless $N{\times}N$ matrix ${\bm{B}}$ is symmetric and tridiagonal,
\begin{equation}
\bm{B} = \frac{m}{K}
\begin{bmatrix}
	\frac{K_{01}{+}K_{12}}{m_1} &  -\frac{K_{12}}{\sqrt{m_1m_2}} & 0 & \hdots & \\[2mm]
	-\frac{K_{12}}{\sqrt{m_1m_2}} & \frac{K_{12}{+}K_{23}}{m_2}
	& -\frac{K_{23}}{\sqrt{m_2m_3}} & & \\
	0 & -\frac{K_{23}}{\sqrt{m_2m_3}} & \frac{K_{23}{+}K_{34}}{m_3}
	& \ddots & \\
	\vdots &     &\ddots&\ddots&   \\
\end{bmatrix}_{N}~,
\label{e.BN}
\end{equation}
and its elements are
\begin{eqnarray}
B_{ii} &=& \frac{m}{K}~\frac{K_{i-1,i}{+}K_{i,i{+}1}}{m_i} ~,
\notag\\
B_{i,i{+}1} &=& -\frac{m}{K}~
\frac{K_{i,i{+}1}}{\sqrt{m_im_{i{+}1}}}  = B_{i{+}1,i}~.
\end{eqnarray}

The chain is assumed to be {\it mirror symmetric}, i.e., the transformation $i\,{\to}\,N{+}1{-}i$ is a symmetry,
\begin{equation}
 K_{i,i{+}1} = K_{N-i,N-i+1}
~,~~~
 m_i = m_{N{+}1-i} ~,
\end{equation}
or $B_{ij}=B_{N{+}1-i,N{+}1-j}$, and, more importantly, the bulk of the chain is assumed to be {\it uniform}, i.e., the elastic constants and the masses are all equal to the typical values, except for a few ones at the endpoints: In particular, only the first (and last) two masses, $m_1=m_N$ and $m_2=m_{N-1}$, as well as the springs attached to $m_1$ and $m_N$, $K_{01}=K_{N,N+1}$ and $K_{12}=K_{N-1,N}$, can differ from the typical values, while
\begin{eqnarray}
 m_3 = m_4 = \dots  = m_{N-2} &=& m ~,
\notag\\
 K_{23} = K_{34} = \dots = K_{N-2,N-1} &=& K ~.
\end{eqnarray}
The bulk elements of $\bm{B}$ are then $B_{ii}=2$ and $B_{i,i\pm{1}}=-1$; it is more convenient to deal with the matrix $\bm{A}\equiv{2}-\bm{B}$, which has a vanishing bulk diagonal,
\begin{equation}
\bm{A} =
\begin{bmatrix}
~z~ &  x  &      &      &      &      &     &     \\
 x  & ~0~ &  y   &      &      &      &     &     \\
	&  y  & ~0~  &  1   &      &      &     &     \\
	&     &  1   & ~0~  &  ~1~ &      &     &     \\
	&     &      &\ddots&\ddots&\ddots&     &     \\
	&     &      &      &  ~1~ & ~0~  &  y  &     \\
	&     &      &      &      &  y   & ~0~ &  x  \\
	&     &      &      &      &      &  x  & ~z~ \\
\end{bmatrix}_{N} ~~.
\label{e.AN}
\end{equation}
The advantage of considering an almost-uniform chain is that the diagonalization of the {\em quasiuniform} matrix $\bm{A}(x,y,z)$ can be analytically afforded~\cite{ABCVV2012,BV2013}.

When writing the matrix $\bm{A}$ in the form~\eqref{e.AN} a further constraint has been implicitly imposed, namely that $A_{22}$ be vanishing: Indeed, the four nonuniform parameters $m_1$, $m_2$, $K_{01}$, and $K_{12}$ are determined by three dimensionless variables $x$, $y$, and $z$. By comparing with the matrix elements $B_{23}$, $B_{22}$, $B_{12}$, and $B_{11}$ in Eq.~\eqref{e.BN} one has
\begin{equation}
\begin{aligned}
 y &= \sqrt{\frac{m}{m_2}}  
\\
 x &= \frac{m}{K}~\frac{K_{12}}{\sqrt{m_1m_2}}
\end{aligned}
~~~~~~
\begin{aligned}
  2 &= \frac{m}{K}~\frac{K_{12}+K}{m_2} 
\\
  z &= 2-\frac{m}{K}~\frac{K_{01}+K_{12}}{m_1} ~,
\label{e.xyz}
\end{aligned}
\end{equation}
the second equality expressing the constraint mentioned above.
There will be no ambiguity if, from now on, one assumes the typical values as measure units (equivalent to setting $m=1$ and $K=1$), so the time unit is $\sqrt{m/K}$. The physical parameters can be expressed in terms of the variables $(x,y,z)$:
\begin{equation}
\begin{aligned}
 m_2&=\frac1{y^2} 
\\
 m_1 &= \frac{(2{-}y^2)^2}{x^2y^2}
\end{aligned}
~~~~
\begin{aligned}
 K_{12} &= \frac{2{-}y^2}{y^2} = 2m_2-1
\\
 K_{01} &= \frac{2{-}y^2}{x^2y^2}~\big[(2{-}y^2)(2{-}z)-x^2\big] ~.
\end{aligned}
\label{e.mKxyz}
\end{equation}
Since these physical parameters have to be positive, the possible values of the variables have constraints:
\begin{equation}
\begin{aligned}
 0 < y < \sqrt2  ~,~~~~ x > 0 ~,
\\
 z \le z_0 \equiv 2-\frac{x^2}{2{-}y^2} ~.
\end{aligned}
\end{equation}
Note that if $x^2>2\,(2{-}y^2)$, then $z$ has to be negative. When $z\,{=}\,z_0$ the chain-wall elastic constant $K_{01}$ vanishes and the chain is {\em isolated} or {\em free}. In order that all elastic constants be equal, it is necessary that $y\,{=}\,1$ and that $z=2(1{-}x^2)$.

The chain has two different {\em uniform limits}, both with equal masses and equal internal elastic constants:
\begin{itemize}
 \item[] {\,$(x,y,z)=(1,1,1)$\,} ~~uniform-free (uf),
 \item[] {\,$(x,y,z)=(1,1,0)$\,} ~~uniform-bounded (ub),
\end{itemize}
namely, the chain without or with the spring connecting to the external walls,  $K_{01}\,{=}\,0$ or $1$; in the first case the chain has to have a zero-frequency (Goldstone) mode.

Let $\bm{U}\,{=}\,\{U_{ni}\}$ be the orthogonal matrix that
diagonalizes $\bm{A}$ (and, of course, also $\bm{B}=2-\bm{A}$),
\begin{equation}
 \sum_{ij}U_{ni}A_{ij}U_{mj} = \lambda_n\,\delta_{nm}~.
\label{e.Uni}
\end{equation}
Then, the diagonal form of the Hamiltonian~\eqref{e.H} reads
\begin{equation}
\H=\frac12\sum_{n=1}^N \big(p_n^2 + \omega_n^2\,q_n^2\big) ~,
\end{equation}
with the eigenfrequencies
\begin{equation}
 \omega_n  =\sqrt{2{-}\lambda_n}
\label{e.omegan}
\end{equation}
and the normal-mode coordinates and momenta
\begin{equation}
 q_n=\sum_{i=1}^NU_{ni}\,q_i
~,~~~
 p_n=\sum_{i=1}^NU_{ni}\,p_i ~.
\label{e.pnqn}
\end{equation}

For given initial conditions $\bm{q}(0)=\bm{x}$ and $\dot{\bm{q}}(0)=\bm{v}$, the  chain dynamics is a superposition of the motions of the normal-modes,
\begin{equation}
 q_i(t) =  \sum_{n=1}^N U_{ni}\,\Big(x_n\cos\omega_nt +\frac{v_n}{\omega_n}\sin\omega_nt\Big)~
\label{e.solution}
\end{equation}
where $x_n$ and $v_n$ are the normal-mode initial values obtained as in Eq.~\eqref{e.pnqn}.

\subsection{Transmission amplitude}
\label{ss.amplitude}

The goal is to start from a static configuration where only the first mass is displaced, $\bm{q}(0)=(x_1,\,0,\,0,\dots,\,0)$, and to look for values of the chain parameters~\eqref{e.mKxyz} such that the dynamics leads in a certain time $t^*$ as close as possible to the mirror-symmetric configuration  $\bm{q}(t^*)=(0,\,0,\,0,\dots,\,x_1)$. With these initial conditions Eq.~\eqref{e.solution} becomes
\begin{equation}
  q_i(t) = x_1\sum_{n=1}^NU_{ni}U_{n1}\cos\omega_nt ~.
\label{e.qit}
\end{equation}
Now, one can use an interesting property~\cite{CantoniB1976} arising from the mirror symmetry of the matrix $\bm{A}$: Assuming that the eigenvalues $\{\lambda_1,\,\lambda_2,\,\dots,\,\lambda_N\}$ are chosen  in decreasing order~\cite{note-omegaincr}, it tells that $U_{nN}\,{=}\,(-)^{n-1}U_{n1}=U_{n1}\,\cos[\pi(n{-}1)]$. This gives, for the last mass in the array,
\begin{equation}
 q_{_N}(t) = x_1~\alpha_{_N}(t)
\end{equation}
where the {\em transmission amplitude}
\begin{equation}
 \alpha_{_N}(t) \equiv  \sum_{n=1}^N\P_n\cos[\pi(n{-}1){-}\omega_nt]
\label{e.utn}
\end{equation}
has been defined. The positive numbers
\begin{equation}
 \P_n \equiv U_{n1}^2 ~,
\label{e.Pn}
\end{equation}
satisfying $\sum_n\P_n\,{=}\,1$ by the orthogonality of $\bm{U}$, can be considered a probability distribution that can be dubbed the {\em mode density}, since they weigh the contribution to $\alpha_{_N}(t)$ from each normal mode. The transmission amplitude has the form of the average over the  mode density of time-dependent phase factors. The difference $\delta_{_N}(t)=1\,{-}\,\alpha_{_N}(t)$ will be dubbed the transmission {\em loss}. Zero loss corresponds to perfect transmission and can only occur at some time instant $t^*$ if all phases are coherent, i.e., they are equal or differ by integer multiples of $2\pi$. For instance, if the eigenfrequencies were equally spaced, say $\omega_n=\delta\omega\,(n{-}1)$, one would have
\begin{equation}
 \alpha_{_N}(t)=\sum_{n=1}^N\P_n\,\cos[(n{-}1)(\pi-\delta\omega\,t)] ~,
\label{e.utNperfect}
\end{equation}
and at the time $t^*=\pi/\delta\omega$ (and odd multiples of it) there would be perfect response, $\alpha_{_N}(t^*)=1$, i.e., the starting elongation would be fully reproduced at the time $t^*$ in the opposite end of the chain. Such ideal behavior is however almost impossible in any discrete array, as the frequencies are not equally spaced and the normal modes will not superpose coherently: This is the phenomenon of {\em dispersion}. Nevertheless, the parameters $(x,y,z)$ can be tuned in such a way as to get $\alpha_{_N}(t^*_{_N})$ quite close to 1 at some time $t^*_{_N}$, as it will be shown in the following. In this regime the overall dynamics described by Eq.~\eqref{e.qit} appears as the formation of a localized wave packet that travels at constant velocity along the chain and gives rise to the elongation of the last mass; in the further evolution the wave packet comes back to the first mass, and so on, like in a Newton cradle.

One can also imagine a different initial situation, starting from the equilibrium configuration $\bm{q}(0)=\bm{0}$, but with a momentum given to the first mass, e.g., by an instantaneous collision, so that $\dot{\bm{q}}(0)=(v_1,\,0,\,\dots,\,0)$, which is the case experimentally studied in Ref.~\onlinecite{RammPH2014}. The dynamics is wanted to lead to  $\dot{\bm{q}}(t^*)=(0,\,\dots,\,0,\,v_1)$, meaning that the same momentum is present at the opposite end. In this case, Eq.~\eqref{e.solution} yields
\begin{equation}
 \dot q_{_i}(t) = v_1\sum_{n=1}^NU_{ni}U_{n1}\cos\omega_nt ~,
\label{e.dotqit}
\end{equation}
and it follows that the relevant ratio $\dot{q}_{_N}(t)/v_1=\alpha_{_N}(t)$ between the transmitted and the initial momentum is given by the same transmission amplitude~\eqref{e.utn}.

If the chain is free, $K_{01}=0$, then by momentum conservation the chain's center of mass uniformly translates with velocity $V=P_1/M$, with $M=\sum_im_i$ and $P_1=v_1\sqrt{m_1}$; if transmission is perfect, then the elastic energy is again zero at time $t^*$, so all spacings must be preserved, and the final configuration is shifted by $\delta{Q}=Vt^*$. In Eq.~\eqref{e.solution} this translation is accounted for by the zero-frequency mode.

Figure~\ref{f.cradle4} shows a possible realization of a mass-spring cradle, involving two auxiliary masses that periodically transmit/receive momentum by hard-sphere collision with the chain extrema; the mass on the right is a distance $\delta{Q}$ apart. In the case of $N\,{=}\,4$ masses transmission can be perfect, as shown in the next subsection. The following sections are devoted to the maximization of the transmission amplitude~\eqref{e.utn} by optimizing the values of the parameters $(x,y,z)$. These optimal values correspond to the extremal masses and springs that yield almost-perfect transmission also for large $N$.

\subsection{Perfect transmission for small $N$}
\label{ss.N3N4}

Within the model~\eqref{e.xyz} the chain can yield ideal transmission if the masses are less than five. Indeed, for isolated chains ($K_{01}{=}0$) with three or four masses the free parameters can be set such as to yield equally spaced frequencies~\cite{HermannS1981}. The same result can be obtained for $N\,{=}\,5$ by allowing for a third free parameter, i.e., releasing the constraint $A_{22}\,{=}\,0$, equivalent to the second of the relations~\eqref{e.mKxyz}. Longer chains can be made ``perfect'' by allowing for more parameters: Of course, such ``engineered'' chains are not uniform. Note that for perfect chains any initial configuration evolves to itself after the period $2\pi/\delta\omega$, where $\delta\omega$ is the frequency spacing; if the chain is mirror symmetric, then after a half period $\pi/\delta\omega$ the perfectly reflected configuration is attained. 

\medskip

For $N=3$ the three masses are $(m_1,1,m_1)$ and the two spring constants are $(1, 1)$; there is only one adjustable parameter, $m_1$. The interaction matrix is
\begin{equation}
\bm{B} =
\begin{bmatrix}
	x^2 & -x  & ~0  \\
	-x  &  2  & -x  \\
	~0   & -x  & x^2 
\end{bmatrix}~~,
\label{e.N3}
\end{equation}
with $x=m_1^{-1/2}$ corresponding to the notation~\eqref{e.mKxyz} with $y\,{=}\,1$. The eigenvalues of $\bm{B}$ are solutions of
\begin{equation}
 \det(\mu-\bm{B}) = \mu\,(\mu{-}x^2)\,(\mu-2{-}x^2) = 0 ~,
\end{equation}
so the eigenvalues are $\mu_1\,{=}\,0$,  $\mu_2\,{=}\,x^2$,  $\mu_3\,{=}\,2{+}x^2$, and the eigenfrequencies $\omega_n=\sqrt{\mu_n}$ are equally spaced if $\omega_3\,{=}\,2\omega_2$, i.e., $2{+}x^2\,{=}\,4x^2$, that has the solution $x^2\,{=}\,2/3$. So three masses with two identical springs show perfect transmission when
\begin{equation}
 m_1=x^{-2}=\fr{3}{2} ~,
\end{equation}
namely the mass sequence is $\propto(3,2,3)$. The frequency spacing is $\omega_1=\sqrt{2/3}$ and the arrival time is $t_3=\pi/\omega_1=\sqrt{3/2}\,\pi\simeq{3.848}$.

\medskip

For $N=4$ the sequence of masses is $(m_1, m_2, m_2, m_1)$, connected by the spring constants $(K_{12},1,K_{12})$, so the interaction matrix is
\begin{equation}
\bm{B} =
\begin{bmatrix}
	~w~ & -x   &  0   &  0  \\
	-x  &  2   & -y^2 &  0  \\
	0  & -y^2 &  2   & -x  \\
	0  &  0   & -x   &  w 
\end{bmatrix}~~,
\label{e.N4}
\end{equation}
where $w\equiv{x^2}/(2{-}y^2)$; it is convenient to set $r\equiv2{-}y^2$, so $x^2=wr$. The secular equation, $0\,{=}\,\det(\lambda\,{-}\,\bm{B})$, is easily worked out and reads
\begin{equation}
\begin{aligned}
 0 &=  [(\mu{-}w)(\mu{-}2)-x^2]^2-(\mu{-}w)^2y^4
\\
   &= \mu\,[\mu-(w{+}r)]
      [\mu^2-\mu(w{+}4{-}r)+2w(2{-}r)]~;
\end{aligned}
\end{equation}
the eigenvalues of $\bm{B}$ are given, in the increasing order $\mu_1<\mu_2<\mu_3<\mu_4$, by
\begin{equation}
\begin{aligned}
 \mu_1 &= 0
 \\
 \mu_3 &= w{+}r
\end{aligned}
~~~~~~~~
\begin{aligned}
 \mu_2+\mu_4 &= w{+}4{-}r
 \\
 \mu_2\,\mu_4 &= 2w(2{-}r) ~~.
\end{aligned}
\end{equation}
In order that the frequencies $\omega_n=\sqrt{\mu_n}$ be equally spaced one has to require $\mu_n=(n{-}1)^2\mu_2$, giving
\begin{equation}
 \mu_4+\mu_2 = \fr{5}{2}\,\mu_3
~,~~~~~~~~~
 \mu_4\,\mu_2= \fr{9}{16}\,\mu_3^2 ~,
\end{equation}
which in terms of the parameters $r$ and $w$ reduce to the linear system
\begin{equation}
 3w+7r=8
~,~~~~~
 2w+r=2 ~.
\end{equation}
The solution is $r\,{=}\,{10}/{11}$ and $w\,{=}\,{6}/{11}$, hence $y^2\,{=}\,{12}/{11}$ and $x\,{=}\,{\sqrt{60}}/{11}$, so the 4-mass perfectly transmitting chain must have
\begin{equation}
 m_1=\fr{55}{36}
~,~~~
 m_2=\fr{11}{12}
~,~~~
 K_{12}=\fr{5}{6} ~.
\end{equation}
Basically, the mass sequence has to be $\propto(5,3,3,5)$ and the spring sequence $\propto(5,6,5)$. The frequency spacing is $\omega_1=\sqrt{w{+}r}/2={2}/{\sqrt{11}}$, which entails the arrival time $t_4=\pi/{\omega_1}={\sqrt{11}\pi}/2\simeq{5.210}$.

\medskip
Also for $N=5$ a perfect solution is known:
\begin{equation}
 m_1=\fr {35}{18}
~,~~~
 m_2=\fr {10}{9}
~,~~~
 K_{12}=\fr {7}{9} ~.
\label{e.N5}
\end{equation}
It is reported in Ref.~\onlinecite{HermannS1981} without proof (here given in Appendix~\ref{a.N5}); the frequencies are $\omega_n=(n{-}1)/{\sqrt{5}}$ for $n=1,...,5$, and entail the arrival time $t_5=\sqrt{5}\pi\simeq{7.025}$. However, this solution does not belong to the constrained model~\eqref{e.xyz} considered in this paper in order to deal with the quasiuniform matrix~\eqref{e.AN}.

\section{Analytic approach}
\label{s.algebra}

\subsection{Characteristic polynomial and phase shifts}

The $N$th-degree characteristic polynomial associated with the $N{\times}N$ matrix $\bm{A}$ [Eq.~\eqref{e.AN}] is given by
\begin{equation}
 \chi_{_N}(\lambda;x,y,z) \equiv \det [\lambda-\bm{A}(x,y,z)] ~;
\label{e.chiN}
\end{equation}
by expanding it in the last column, one finds
\begin{equation}
 \chi_{_N}= [(\lambda{-}z)\lambda-x^2]~\xi_{_{N-2}}-(\lambda{-}z)y^2~\xi_{_{N-3}}~,
\label{e.chixi}
\end{equation}
where $\xi_{_N}$ is the characteristic polynomial of the matrix with only one nonuniform endpoint,
\begin{equation}
\xi_{_N}(\lambda;x,z) =
\begin{vmatrix}
	\lambda-z & -x     &        &        &        &         \\
	-x       &\lambda &   -y   &        &        &         \\
	& -y     &\lambda &   -1   &        &         \\
	&        & \ddots & \ddots & \ddots &         \\
	&        &        &  -1    &\lambda & -1      \\
	&        &        &        & -1     & \lambda \\
\end{vmatrix}_{N}~.
\end{equation}
The same kind of  expansion in the first column gives a similar relation,
\begin{equation}
 \xi_{_N}= [(\lambda{-}z)\lambda-x^2]~\eta_{_{N-2}}-(\lambda{-}z)y^2~\eta_{_{N-3}} ~,
\label{e.xieta}
\end{equation}
in terms of the characteristic polynomials
\begin{equation}
 \eta_{_N}(\lambda)\equiv\chi_{_N}(\lambda;1,1,0)
\label{e.uniformbounded}
\end{equation}
of the fully uniform matrices $A_{_N}(1,1,0)$, which correspond to the uniform-bounded chain. These polynomials are well-known in the literature, as their roots can be obtained in a simple way: By column expansion one has the recursion relation
\begin{equation}
 \eta_{_N}= \lambda\,\eta_{_{N-1}}-\,\eta_{_{N-2}} ~,
\label{e.etaeta}
\end{equation}
which, using the initial conditions $\eta_{_0}=1$ and $\eta_{_1}=\lambda$, can be solved in terms of Chebyshev polynomials of the second kind: Using, in the place of $\lambda$, the real variable $k\in[0,\pi]$ given by
\begin{equation}
 \lambda \equiv 2\cos k = e^{ik}+e^{-ik}
 \label{e.lambda-k}
\end{equation}
they can be synthetically expressed as
\begin{equation}
 \eta_{_N}(\lambda) = \frac{\sin(N{+}1)k}{\sin{k}}
   =\frac{\Im\big\{e^{i(N{+}1)k}\big\}}{\sin{k}} ~.
\label{e.chebyshev}
\end{equation}

Hence, for the uniform-bounded chain, that has the secular equation $\eta_{_N}(\lambda)=0$, the $N$ eigenvalues correspond to the following values of $k$:
\begin{equation}
 k_n^{\rm(ub)} = \frac{\pi\,n}{N{+}1}~,~~~~~~ (n=1,\dots , N)~.
\label{e.k_x1y1}
\end{equation}

Note that, keeping the notation~\eqref{e.lambda-k}, in terms of the pseudo wavevector $k$ the system frequencies~\eqref{e.omegan} read
\begin{equation}
		\omega_k = \sqrt{2-2\cos{k}}=2\,\sin\textstyle\frac{k}2~.
\label{e.omegak}
\end{equation}
For $k\,{\ll}\,1$ (i.e., $n\,{\ll}\,N$) they are almost linear in $k$ and equally spaced in $n$, i.e., $\omega_k\simeq{k}\propto{n}$, meaning that wave packets with main components in the small-$k$ zone travel with low dispersion along the chain. This corresponds to the continuum limit, namely the elastic string.

In order to work out Eq.~\eqref{e.chixi} one first uses Eq.~\eqref{e.xieta}
\[
\begin{aligned}
 \chi_{_N}=~&(\lambda^2{-}z\lambda{-}x^2)
 \big[(\lambda^2{-}z\lambda{-}x^2)\eta_{_{N-4}}-(\lambda{-}z)y^2\eta_{_{N-5}}\big]
\\
 &~-(\lambda{-}z)y^2\big[(\lambda^2{-}z\lambda{-}x^2)\eta_{_{N-5}}
   -(\lambda{-}z)y^2\eta_{_{N-6}}\big]~,
\end{aligned}
\]
and then by Eq.~\eqref{e.chebyshev},
\begin{equation}
 \sin{k}~\chi_{_N}(k) = \Im\big\{e^{i(N{+}1)k} u_k^2\big\} ~,
\label{e.chik}
\end{equation}
where
\begin{equation}
 u_k\, ~\equiv~
 e^{-2ik}\big[(\lambda^2{-}z\lambda{-}x^2)-(\lambda{-}z)y^2e^{-ik}\big]
\label{e.uk1}
\end{equation}
is a complex variable whose dependence on $k$ can be made fully explicit using Eq.~\eqref{e.lambda-k},
\begin{equation}
\begin{aligned}
 u_k = ~& 1-ze^{-ik}+(2{-}x^2{-}y^2)e^{-2ik}
\\
  &~-z(1{-}y^2)e^{-3ik}+(1{-}y^2)e^{-4ik} ~.
\label{e.uk}
\end{aligned}
\end{equation}
Equation~\eqref{e.chik} yields a compact expression of the secular equation, $\chi_{_N}(k)=0$. For instance, in the uniform-bounded case it is $u_k=1$, so its solution~\eqref{e.k_x1y1} is immediate. Setting
\begin{equation}
 u_k \equiv |u_k|\,e^{-i\varphi_k}~,
\label{e.ukpolar}
\end{equation}
the secular equation reads $\Im\big\{e^{i[(N{+}1)k-2\varphi_k]}\big\}=0$, so that the
eigenvalues $\lambda_n\,{=}\,2\cos{k_n}$ correspond to those values of $k\,{=}\,k_n$ such that $(N{+}1)k-2\varphi_k=\pi\,n$ and are given as corrections to the uniform-bounded values, Eq.~\eqref{e.k_x1y1}, due to the {\em phase shifts}~ $\varphi_k$,
\begin{equation}
 k_n = \frac{\pi\,n+2\varphi_{k_n}}{N{+}1}  ~,~~~~~~ (n=1, \dots , N)~.
\label{e.kn}
\end{equation}
By the way, this determines the sign of the real quantity in braces in Eq.~\eqref{e.chik},
\begin{equation}
 e^{i(N{+}1)k_n}u_{k_n}^2=e^{i\pi{n}}|u_{k_n}|^2=(-)^n\,|u_{k_n}|^2 ~.
\end{equation}

Note that there is a one-to-one correspondence between the $N$ allowed values $\{k_n\}$ and the indices $n=1,\dots,N$, so it is unambiguous to use the index $k$ in the place of $n$, as mostly done in the following.

\subsection{The mode density has to be peaked at $k=0$}

The mode density~\eqref{e.Pn} is calculated in Appendix~\ref{a.modedensity}, where it is shown that a compact expression can be worked out, namely:
\begin{equation}
 \P_k= \frac{2x^2y^2\sin^2\!{k}}{(N{+}1{-}2\varphi_k')\,|u_k|^2}~.
\label{e.Uk1b}
\end{equation}
In the uniform-bounded limit, where $u_k\,{\to}\,1$ and $\varphi_k\,{\to}\,0$, this agrees with the known result.

After the discussion of Sec.~\ref{ss.amplitude}, it is clear that a high transmission amplitude can be obtained only if the normal modes selected by $\P_k$, lie in a zone where the frequencies~\eqref{e.omegak} are almost equally spaced. This happens in the ``linear region'' around $k\,{=}\,0$, meaning that the parameters $(x,y,z)$ have to be chosen such that $\P_k$ be peaked around $k\,{=}\,0$. Due to the sine function in Eq.~\eqref{e.Uk1b}, for $\P_{k=0}$ to be at least finite it must be $u_{k=0}=0$, i.e.,
\begin{equation}
 1-z+(2{-}x^2{-}y^2)-z(1{-}y^2)+ (1{-}y^2)= 0 ~.
\end{equation}
To satisfy this it is natural to keep $x$ and $y$ as free parameters, while fixing the value of $z$,
\begin{equation}
  z(x,y) \equiv 2- \frac{x^2}{2{-}y^2} ~.
\label{e.z0}
\end{equation}
Note that the second of Eqs.~\eqref{e.mKxyz} tells that this implies $K_{01}\,{=}\,0$\,: The chain ends do not interact with the walls. In other words, it has been found that a requisite for coherent pulse propagation is that the spring-mass chain be isolated. From now on the condition~\eqref{e.z0} is assumed to hold.

As~\eqref{e.uk} shows that $u_k$ is a polynomial in $e^{-ik}$, one can extract from $u_k$ the factor $(1{-}e^{-ik})$, obtaining
\begin{equation}
 u_k = (1{-}e^{-ik})~\tu_k
 ~\equiv~ 2\,\sin\fr{k}2~e^{-i(k{-}\pi)/2}~\tu_k ~,
\label{e.ukz0}
\end{equation}
with
\begin{equation}
 \tu_k \equiv 1-(1{-}w)e^{-ik}{-}(1{-}w)(y^2{-}1)e^{-2ik}{+}(y^2{-}1)e^{-3ik}
\label{e.tudef}
\end{equation}
where the new parameter
\begin{equation}
 w \equiv \frac{x^2}{2{-}y^2} ~,
\label{e.w}
\end{equation}
here introduced for brevity, will be useful in the following.

\subsection{Phase shifts for the free chain}
\label{ss.shifts}

The case $x\,{=}\,y\,{=}1$ with the choice~\eqref{e.z0} is to be understood as the uniform-free limit, since $z(1,1){=}1$. This means that $\tu_k\,{=}\,1$ and Eq.~\eqref{e.ukz0} tells that the phase shifts~\eqref{e.ukpolar} are $\varphi_k\,{=}\,(k{-}\pi)/2$. From Eq.~\eqref{e.kn} the allowed values of $k$ take the equally spaced values
\begin{equation}
 k_n^{\rm (uf)}=\frac{\pi(n{-}1)}N~,~~~~~~(n=1,\dots,N)~;
\label{e.knfree}
\end{equation}
these include the Goldstone mode $k_1\,{=}\,0$ with frequency $\omega_1\,{=}\,0$, i.e., the translation mode expected for the isolated chain. As observed in the previous subsection, being $\omega_k\simeq{k}$ for $k\ll{1}$, a situation close to that of perfect transmission, Eq.~\eqref{e.utNperfect}, would arise if the mode density selected only low-$k$ modes: Indeed, in Eq.~\eqref{e.utn} the phases would be
\begin{equation}
 \pi(n{-}1){-}2t\sin k_n \simeq \frac{\pi(n{-}1)}N\,(N{-}t) ~.
\end{equation}
Actually this is not the case, since in the uniform-free case the density~\eqref{e.Uk1b} is peaked in zero, indeed, but consists in a too broad distribution~\cite{note-P0half},
\begin{equation}
 \P_k^{\rm (uf)} = \frac2N~\cos^2\!\frac{k}2~;
\label{e.Pkunif}
\end{equation}
the purpose is to vary the parameters $x$ and $y$ in such a way to deform this density and make it narrower; however, shrinking $\P_k$ too much would definitely be disadvantageous, because the phase shifts change with $x$ and $y$ and the spacings between eigenvalues are deformed by Eq.~\eqref{e.kn}. Therefore, one expects that an optimal compromise will maximize the transmission amplitude at some arrival time.

The phase $\psi_k$ of 
\begin{equation}
 \tu_k = |\tu_k|~e^{i\psi_k}
\label{e.tupsik}
\end{equation}
turns out to give the phase-shift correction to the uniform-free solution~\eqref{e.knfree}:  Indeed Eq.~\eqref{e.ukz0} gives $\psi_k\equiv(k{-}\pi)/2-\varphi_k$, and  Eq.~\eqref{e.kn} becomes
\begin{equation}
 k_n = \frac{\pi\,(n{-}1)-2\psi_{k_n}}{N} 
~,~~~~~~ (n=1,\dots, N)~.
\label{e.kn1}
\end{equation}
The Goldstone mode $k_1\,{=}\,0$ is preserved, since $\tu_{k=0}$ is real and the corresponding phase shift vanishes, $\psi_{k_1}\,{=}\,0$.

In the following it will appear to be useful to define, besides the variable $w$, Eq.~\eqref{e.w}, a further variable $r$,
\begin{equation}
 \left\{\begin{aligned} r &\equiv 2\,{-}\,y^2 \\ w &\equiv x^2/r \end{aligned}\right.
 ~~~\Longleftrightarrow~~~
 \left\{\begin{aligned} y^2 &= 2\,{-}\,r \\ x^2 &= r\,w \end{aligned}\right. ~,
\label{e.rw}
\end{equation}
so that $w$ and $r$ can be used in the place of $x$ and $y$; indeed, it will be proven later that the optimal values of both $r$ and $w$ decrease as a negative power of $N$. Note that the uniform limit corresponds to $r\,{=}\,w\,{=}\,1$. In Appendix~\ref{a.density_rw} the dependence of the mode density on $w$ and $r$ is worked out and made explicit in Eq.~\eqref{e.Pk}. By means of it and using the frequencies~\eqref{e.omegak} the transmission amplitude Eq.~\eqref{e.utn} is given by 
\begin{equation}
 \alpha_{_N}(t)= \sum_{n=1}^N \P_{k_n}
            ~\cos\big[\pi{(n{-}1)}-2t\,\sin\!\fr{k_n}2\big] ~.
\label{e.utkn}
\end{equation}
where the discrete pseudo wavevectors $\{k_n\}$ are the solutions of Eq.~\eqref{e.kn1}. In the following sections these formulas are used for setting up a numerical approach in order to calculate the optimal values of the parameters $x^{*}$ and $y^{*}$ which yield at some time $t^{*}$ the maximum attainable transmission amplitude $\alpha^*$. In addition the asymptotic behaviors in the limit of large $N$ will be exactly derived.

\section{Optimizing one mass}
\label{s.m1}

In this section the simpler case when $y=1$ (i.e., $r\,{=}\,1$ and $w\,{=}\,x^2$) is considered, which, by Eqs.~\eqref{e.mKxyz}, corresponds to the mass-spring chain with all spring constants and all masses equal, except for the first and the last masses,
\begin{equation}
 m_1 = m_N = \frac{1}{x^2} ~.
\label{e.m1x}
\end{equation}

\subsection{Mode density and phase shifts}

Setting $r\,{=}\,1$ and $w\,{=}\,x^2$ in Eq.~\eqref{e.Pk}, the density becomes a pseudo-Lorentzian~\cite{note-P0half} peaked at $k\,{=}\,0$\,,
\begin{equation}
\P_k =  \frac1{N{+}2\psi_k'}
~\frac{\Delta(1{+}\Delta)}{\Delta^2+\tan^2\!\fr{k}2} ~,
\end{equation} 
with the parameter
\begin{equation}
 \Delta \equiv  \frac{x^2}{2{-}x^2}
\label{e.xDelta}
\end{equation}
characterizing the distribution width. A convenient expression for the phase shifts defined in Eq.~\eqref{e.tupsik} is found using Eq.~\eqref{e.tudef},
\begin{equation}
\begin{aligned}
 \tu_k  &\equiv |\tu_k|~e^{i\psi_k}
               = 1-(1{-}x^2)e^{-ik}
\\
 &= e^{-ik/2}\big[e^{ik/2}-(1{-}x^2)e^{-ik/2}\big]
\\
 &= e^{-ik/2}\big[x^2\cos\fr{k}2+i(2{-}x^2)\sin\fr{k}2 \big] ~,
\end{aligned}
\end{equation}
so
\begin{equation}
 \psi_k =\tan^{-1}\frac{\tan\frac{k}2}\Delta~-\frac{k}2 ~;
\label{e.psiDelta}
\end{equation}
inserting in Eq.~\eqref{e.kn1} the set $\{k_n\}$ corresponding to the normal modes is obtained. Note that $\psi_k$ increases with $k$, which means that the $k_n$'s ($n\,{>}\,1$) get a negative correction with respect to the free case and the frequency spacings decrease. Taking the derivative of $\psi_k$,
\begin{equation}
 2\psi_k' = (1-\Delta)\,\frac{\Delta-\tan^2\fr{k}2}{\Delta^2+\tan^2\fr{k}2}~,
\label{e.psiprime}
\end{equation}
the mode density can be written in fully explicit way,
\begin{equation}
 \P_k = \frac{\Delta(1{+}\Delta)}
		{(N\Delta{+}1{-}\Delta)\Delta+(N{-}1{+}\Delta)\tan^2\!\fr{k}2} ~.
\label{e.PkDelta}
\end{equation}
Its exact normalization is a nontrivial outcome that can be numerically verified, while for $N\,{\to}\,\infty$ this agrees with an analytic result~\cite{Dwight1961}.

\subsection{Frequency spacings and group velocity}

\begin{figure}
\includegraphics[width=0.47\textwidth]{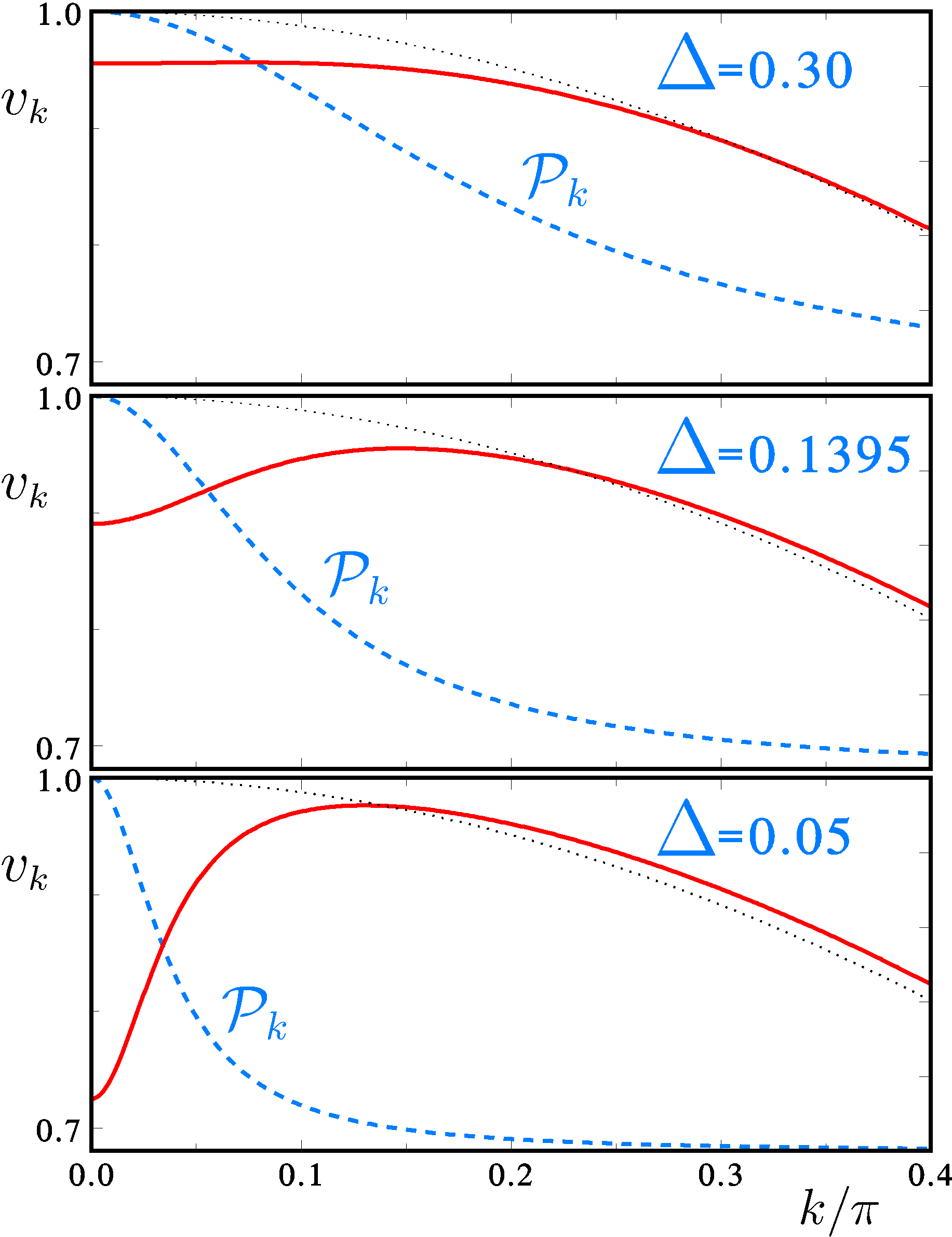}
\caption{The group velocity $v_k$ (red solid line) and the mode density $\P_k$ (blue dashed lines, arbitrary units) for a chain of 50 masses. The three panels correspond to  $\Delta=0.30$, $0.1395$, and $0.05$. For the uniform chain ($\Delta=1$) $v_k$ (dotted line) is initially flat and small-$k$ modes are almost coherent. Decreasing $\Delta$ selects low-$k$ modes, but also deforms $v_k$, destroying coherence. The middle panel shows the optimal compromise given in Table~\ref{t.opt_y1}.}
	\label{f.vk}
\end{figure}

From the expression of $\P_k$ obtained above, the purpose of making $\P_k$ narrower can be easily achieved by decreasing $\Delta$, namely choosing $m_1$ larger than the other masses. However, as mentioned in Sec.~\ref{ss.shifts}, this also affects the phase shifts $\psi_k$ and the spacings between frequencies. In order to appreciate this effect, one is led to look at (an analog of) the group velocity (lattice spacings per time unit),
\begin{equation}
 v_k = \frac{N}\pi\,\partial_n\omega
     = \frac{N}\pi\,\cos\fr{k}2~\partial_n{k} ~,
\label{e.vk}
\end{equation}
which is indeed proportional to the spacing between subsequent values $\omega_n$: The more constant is $v_k$, the more likely is that the initial pulse propagate coherently. In order to highlight the full dependence of $v_k$ on $k$, note that  from~\eqref{e.kn1} one has $(N{+}2\psi'_k)\,\partial_nk=\pi$, so
\begin{equation}
 v_k = \frac{N}{N{+}2\psi_k'}\,\cos\fr{k}2~.
\end{equation}
This expression suggests a strategy to improve the transmission, since one could make $v_k$ flatter by a proper choice of a small $\Delta$; then, Eq.~\eqref{e.psiprime} can be expanded,
\begin{equation}
 2\psi_k'
 = \frac{1{-}\Delta}\Delta-\frac{1{-}\Delta^2}{4\Delta^3}\,k^2 +O(k^4)~,
\end{equation}
and also the group velocity
\begin{equation}
 v_k = \frac{N}{t^*} ~\frac{1{-}\frac18k^2}
       {1-\frac{1{-}\Delta^2}{4\Delta^3t^*}\,k^2}+O(k^4)~.
\label{e.vk}
\end{equation}
Here
\begin{equation}
 t^* \equiv N+\frac{1{-}\Delta}\Delta 
\end{equation}
represents the end-to-end arrival time: Indeed, an excitation traveling with the velocity $v_{k{=}0}=N/t^*$ would cover the distance of $N$ lattice spacings in a time $t^*$; in the uniform limit ~$v_0\,{=}\,1$ and $t^*\,{=}\,N$: The correction to this value is the {\em delay} and is positive since $\Delta\,{<}\,1$. To make $v_k$ flat, an obvious choice is to cancel all quadratic terms in $k$ imposing for $\Delta$ the condition
\begin{equation}
 \frac{1{-}\Delta^2}{4\Delta^3t^*}=\frac18 ~.
\label{e.vgquartic}
\end{equation}
This choice makes the group velocity almost constant in a rather wide $k$-interval, but such a criterion could work only if the involved modes, as weighted by $\P_k$, mainly lie in this flat region. The first panel of Fig.~\ref{f.vk} ($N=50$) shows that this is not the case: While $v_k$ becomes particularly flat for the value that satisfies Eq.~\eqref{e.vgquartic}, $\Delta\simeq0.30$, still the width of $\P_k$ is too large and involves many modes with different group velocities. Therefore, a sharper distribution must perform better, in spite of the more deformed group velocity. Figure~\ref{f.vk} illustrates the conflict between the two effects, shrinking of $\P_k$ and deformation of $v_k$.
What is to be learned here is that one would like to optimize two effects, which is hard when having only one free parameter: Two free parameters are expected to yield better results, as is shown in Sec.~\ref{s.m1m2K12}. In any case, Eq.~\eqref{e.vgquartic} yields the large-$N$ behavior
\begin{equation}
 \Delta\sim 2^{1/3} N^{-1/3}
~,~~~ x\sim 2^{2/3}N^{-1/6} ~,
\end{equation}
a scaling that turns out to be correct, though with different prefactors.

\subsection{Transmission amplitude for large $N$}
\label{ss.x.asymptotic}

The transmission amplitude at the time $t$ is given by Eq.~\eqref{e.utn}; using the frequencies~\eqref{e.omegak}, in the large-$N$ limit one can write the sum as an integral,
\begin{equation}
 \alpha_\infty(t) = \lim_{N\to\infty} \int\limits_1^N dn~
 \P_{k_n}~\cos\big[\pi(n{-}1){-}2t\sin\!\fr{k_n}2\big]~,
\end{equation}
with $k_n$ given by Eq.~\eqref{e.kn1}, which implies
\begin{equation}
 \pi\,dn = (N+2\psi'_k)~dk ~,
\end{equation}
so that
\begin{equation}
 \alpha_\infty(t) = \!\lim_{N\to\infty} \int\limits_0^\pi \!\! \frac{dk}\pi
 \frac{\Delta(1{+}\Delta)}{\Delta^2{+}\tan^2\!\fr{k}2}
 \cos(Nk{+}2\psi_k{-}2t\sin\fr{k}2) \,.
\end{equation}
Writing the arrival time as $t\,{=}\,N\,{+}\,s$, where $s$ is the {\em arrival delay}, and setting $k=2q$,
\begin{equation}
 \alpha_\infty(s) = \frac{2\Delta(1{+}\Delta)}\pi \!
 \int\limits_0^{\pi/2}\!\! dq
 ~\frac{\cos\{2[t(q{-}\sin{q})-sq+\psi_{2q}]\}}{\Delta^2+\tan^2\!q}~.
\end{equation}
As by increasing $N$ the distribution gets narrower and narrower ($\Delta\,{\sim}\,N^{-1/3}$), it is convenient to make the substitution $q=\Delta\,\xi$ to an integration variable $\xi$ of the order of unity and then expand in $\Delta$ keeping all leading terms:
\begin{equation}
\begin{aligned}
 2t\,(q-\sin q) &\simeq {\textstyle\frac13}\,t\,\Delta^3\xi^3 ~,
\\
 \psi_{2q} = \tan^{-1}\frac{\tan q}\Delta-q
 &\simeq~ \tan^{-1}\!\xi ~,
\\
\frac{\Delta~dq}{\Delta^2+\tan^2q}
 &\simeq~ \frac{d\xi}{1+\xi^2} ~.
\end{aligned}
\label{e.leading}
\end{equation}
It is then natural to define rescaled counterparts of the arrival
time $t\simeq{N}$ and of the delay $s\sim{N^{1/3}}$,
\begin{equation}
\tau \equiv \frac{\Delta^3}3~t
~,~~~~~~~
\sigma \equiv 2\Delta~s ~.
\label{e.scaling}
\end{equation}
Optimizing $\Delta$ is now converted into optimizing $\tau$, i.e., the coefficient of the asymptotic scaling law for $\Delta$ vs. $t\simeq{N}$. The final asymptotic expression is
\begin{equation}
 \alpha_\infty(\tau,\sigma)  = \frac{2}\pi \int_0^{\infty}\!\!\! d\xi
	~\frac{\cos(\tau{\xi^3}-\sigma{\xi}+2\tan^{-1}\!\xi)}{1+\xi^2}~.
\label{e.uinfty}
\end{equation}
Using the variable ${\zeta}=\tan^{-1}\!\xi$\,, such that $d\xi=(1{+}\xi^2)~d{\zeta}$, the integral becomes a summation of phase factors,
\begin{equation}
 \alpha_\infty(\tau,\sigma) = \frac{2}\pi \int\limits_0^{\pi/2} d{\zeta}~
	\cos(\tau \tan^3{\zeta}-\sigma\tan{\zeta}+2\zeta) ~,
\label{e.ucos}
\end{equation}
and it is evident that the largest amplitude corresponds to the parameters that minimize the phase, namely the argument of the cosine. The same integral appeared in a different context in Ref.~\onlinecite{ABCVV2011}, where it was evaluated numerically: The maximum attainable amplitude amounts to
\begin{equation}
 \alpha_\infty^{*}=0.846902
\end{equation}
and corresponds to $\sigma^{*}=1.2152$ and $\tau^{*}=0.02483$. From Eq.~\eqref{e.scaling} it follows that
\begin{equation}
 \Delta^{*} \simeq  0.4208~N^{-1/3}
~,~~~~~
 s^{*} \simeq  1.444~N^{1/3}~;
\end{equation}
hence, by Eqs.~\eqref{e.xDelta} and~\eqref{e.m1x}, the variable $x$ and the corresponding extremal mass $m_1$ scale as
\begin{equation}
\begin{aligned}
 x^{*} &\simeq \sqrt{2\Delta} \simeq 0.9173~N^{-1/6}
\\
 m_1^{*} &= \frac1{x^2} \simeq 1.188~N^{1/3} ~.
\end{aligned}
\label{e.asymp_x}
\end{equation}

\subsection{Numerical results for finite $N$}

\begin{figure}
\includegraphics[width=0.47\textwidth]{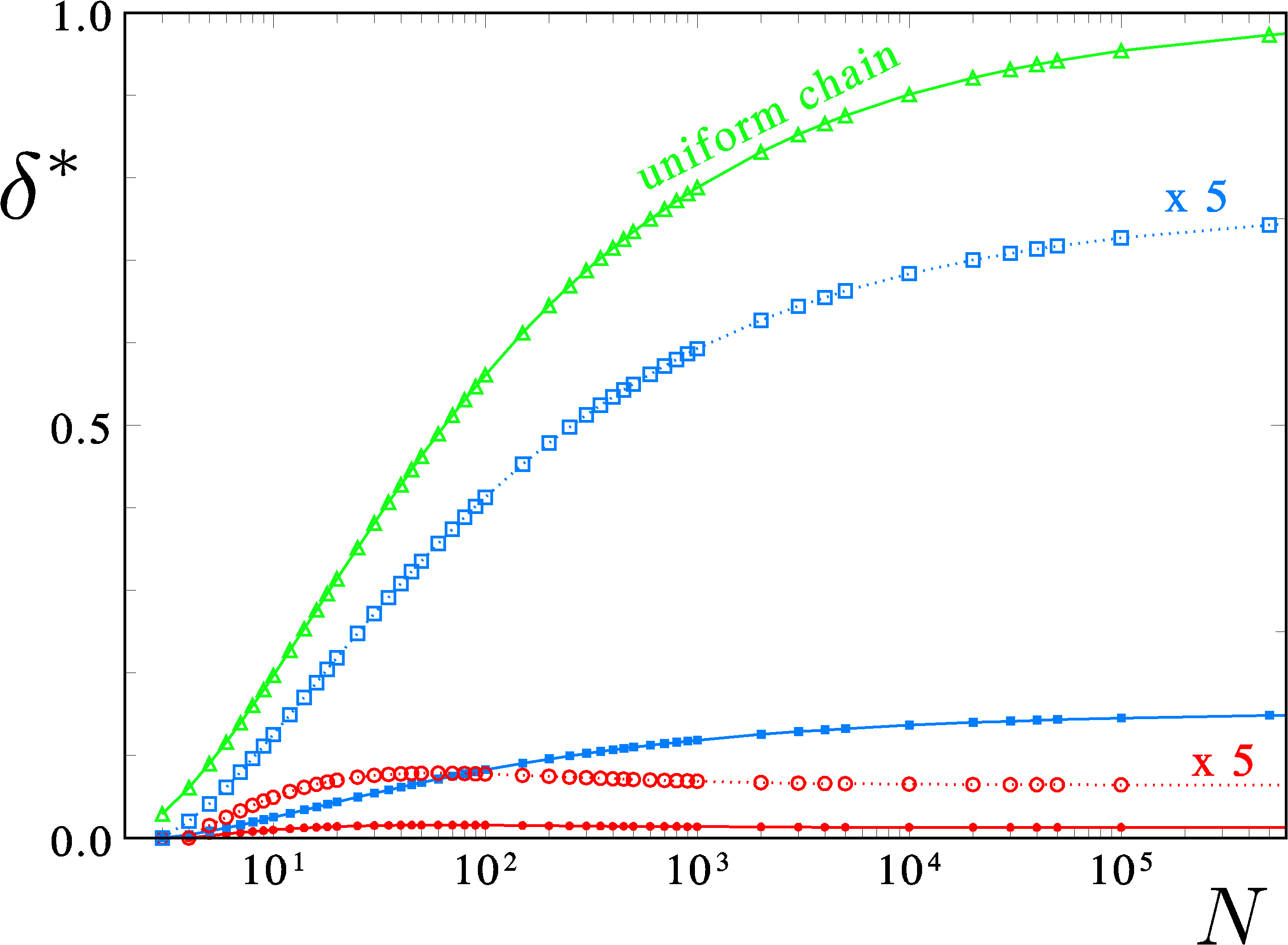}
\caption{Minimal transmission loss $\delta^{*}\equiv\,1{-}\alpha^*$ reported vs. $N$ for the three cases: the uniform chain (green triangles), the chain with optimization of one mass $m_1$ (Sec.~\ref{s.m1}, blue squares), and that with optimization of two masses and their spring,  $m_1$, $m_2$, and $K_{12}$ (Sec.~\ref{s.m1m2K12}, red circles). The asymptotic limits are $1$, $0.1531$, and $0.01285$, respectively. Open symbols are enlargements of the last two cases; the curves are splines. The fully uniform chain is evidently very dispersive, while the optimization of few parameters proves to enormously enhance the quality of transmission: Even in the limit of a (frictionless) infinite chain it is possible to efficiently transmit a pulse.}
	\label{f.delta}
\end{figure}

\begin{figure}
\includegraphics[width=0.47\textwidth]{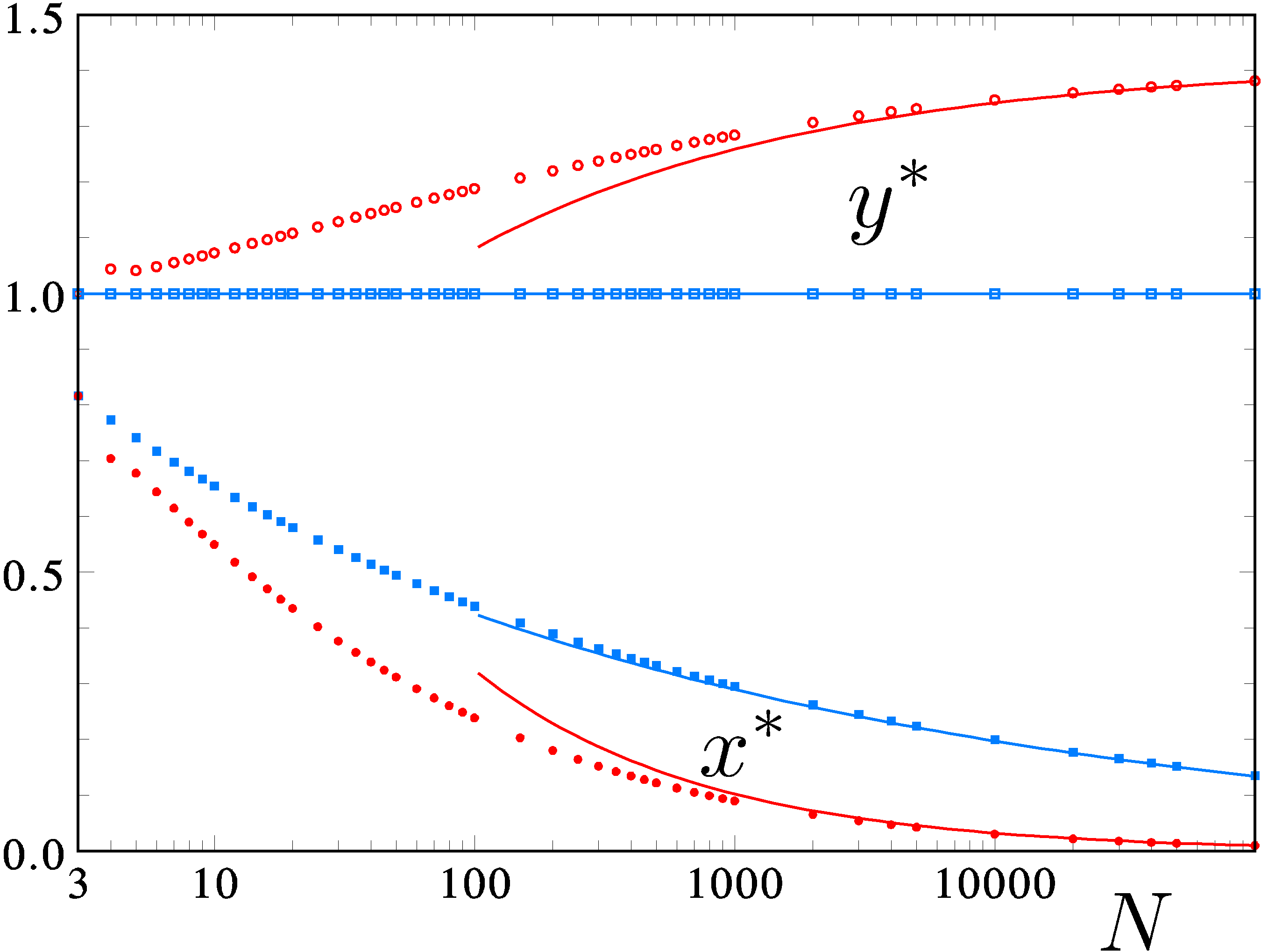}
\caption{Optimal parameters $x^*$ and $y^*$, defined in Eqs.~\eqref{e.mKxyz}, reported vs. $N$ in the case of Sec.~\ref{s.m1}, for which $y=1$ is fixed (blue squares), and the case of Sec.~\ref{s.m1m2K12} (red circles). The curves are the large-$N$ asymptotics of Eqs.~\eqref{e.asymp_x} and~\eqref{e.xy_N}.}
	\label{f.x_xy}
\end{figure}

\begin{figure}
\includegraphics[width=0.47\textwidth]{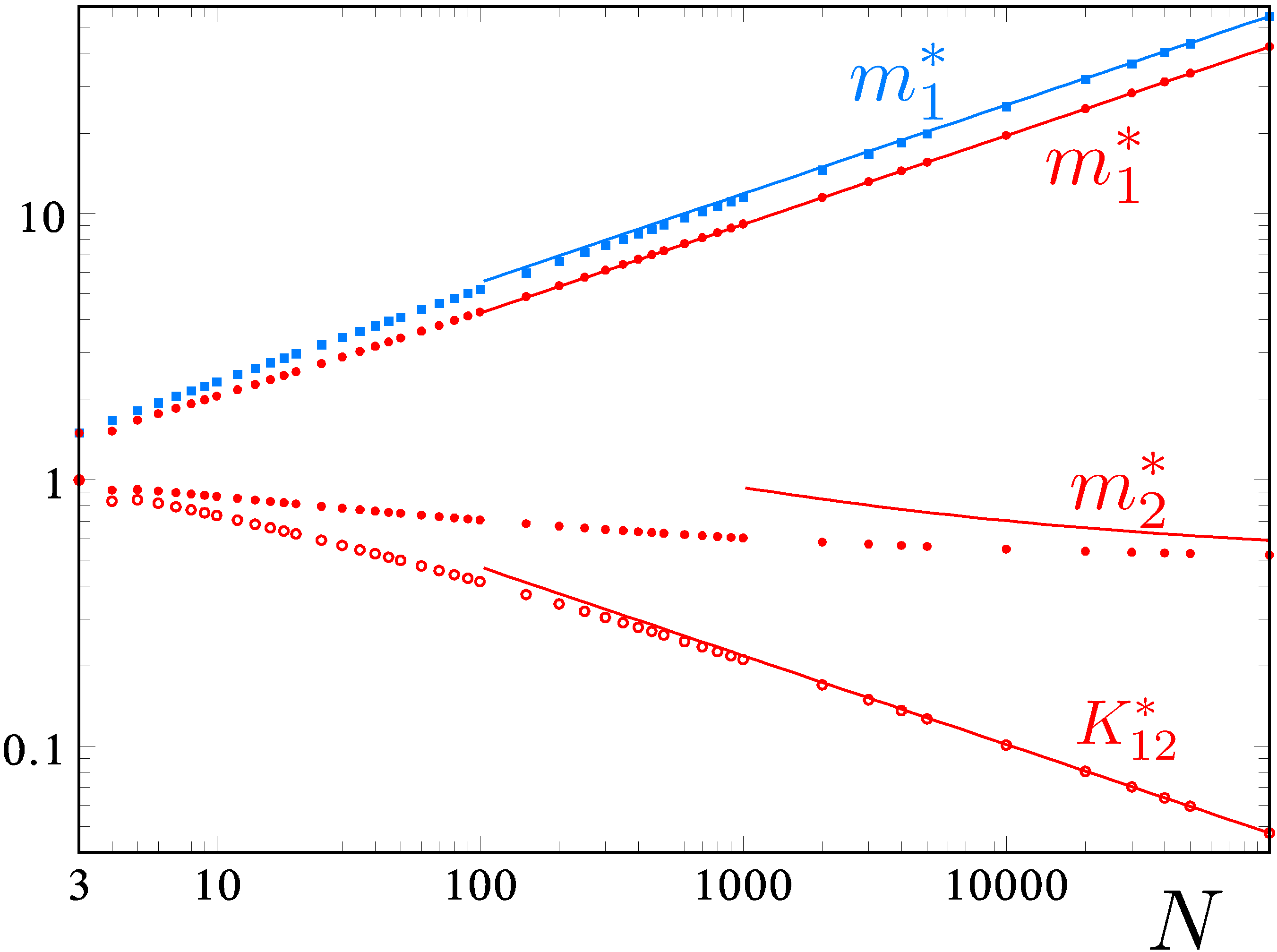}
\caption{Optimal mass- and spring parameters $m_1^*$, $m_2^*$, and $K_{12}^*$, defined in Eq.~\eqref{e.Horig}, are reported for the two cases, namely, the chain of Sec.~\ref{s.m1}, for which $m_2=1$ and $K_{12}=1$ (blue squares), and that of Sec.~\ref{s.m1m2K12} (red circles). The curves are the large-$N$ asymptotics of Eqs.~\eqref{e.asymp_x} and~\eqref{e.m1m2K12-rw}.}
	\label{f.m_mmK}
\end{figure}

The numerical results have been obtained by a code that, for given $N$ and $x$, first evaluates and stores the set $\{k_n\}$ from Eqs.~\eqref{e.kn1} and~\eqref{e.psiDelta} by an iterative algorithm, and then uses the density~\eqref{e.PkDelta} to compute the amplitude $\alpha(s,x)$, Eq.~\eqref{e.utkn}, at different time delays $s\equiv{t}\,{-}\,N$ finding the maximal $\alpha^*[s^*(x),x]$; then an outer loop varies $x$ until finding the best among these maxima, $\alpha^*[s^*(x^*),x^*]$, so identifying the optimal $x^*$ and related delay $s^*(x^*)$. The outcomes for a choice of finite values of $N$ are reported in Table~\ref{t.opt_y1} and illustrated in Figs.~\ref{f.delta}, \ref{f.x_xy} and~\ref{f.m_mmK}. There, the optimal amplitude loss $\delta^*=1-\alpha^*$, which tends to zero in the case of perfect transmission, is reported in the place of $\alpha^*$: The loss increases with $N$ and reaches the asymptotic value $\delta^*_\infty=0.153098$ according to the above calculation. The last column of Table~\ref{t.opt_y1} is the amplitude loss for the fully uniform chain ($x\,{=}\,1$), also shown in Fig.~\ref{f.delta}: It appears that the bare modulation of the extremal masses yields an enormous improvement for any $N$. For instance, in a 20-mass chain the loss can be reduced from 31\% to 4.4\%.

\section{Optimizing two masses and their spring}
\label{s.m1m2K12}

\subsection{Transmission amplitude for large $N$}

\begin{figure}
\includegraphics[width=0.47\textwidth]{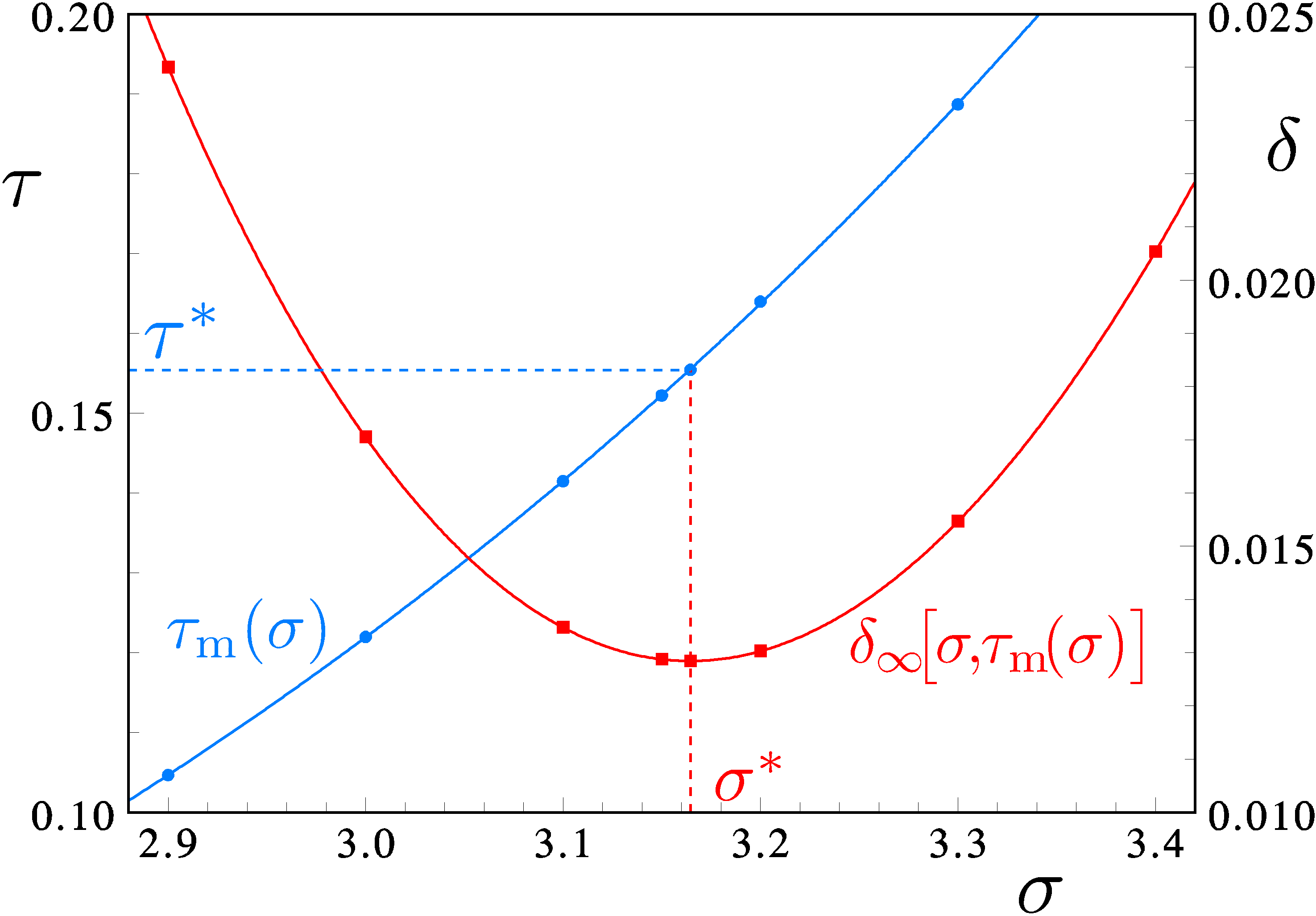}
\caption{Minimization of the asymptotic transmission loss $\delta_\infty(\sigma,\tau)$. First, for different $\sigma$'s one finds the values $\tau_{\rm{m}}(\sigma)$ (blue line, left $y$ axis) that minimize $\delta_\infty\equiv1-\alpha_\infty$ to the value $\delta_\infty\big[\sigma,\tau_{\rm{m}}(\sigma)\big]$ (red line, right $y$ axis); the latter has its global minimum for $\sigma\,{=}\,\sigma^*$.}
\label{f.minimiz}
\end{figure}

It will be justified at the end of this section that to attain maximal transmission, at least in the large-$N$ limit, the denominator of the mode density~\eqref{e.Pk} must be quartic, namely of the type of Eq.~\eqref{e.Pkquartic} below. To satisfy this assumption one has to impose, between the parameters $r$ and $w$, the following constraint:
\begin{equation}
 r^2(2{-}w)^2-32(1{-}r)w = 0~.
\label{e.quarticP}
\end{equation}
Looking for $w(r)$ satisfying Eq.~\eqref{e.quarticP}, one finds
\begin{equation}
 w = 2 \Big(\,\frac{1{-}\sqrt{1{-}r}}{1{+}\sqrt{1{-}r}}~\Big)^2
 = \frac2{r^2} \big(1-\sqrt{1{-}r}\big)^4 ~,
\label{e.wr}  
\end{equation}
the first expression being useful for the numerics, the second for expanding: For small $r$ it gives $w=\frac{r^2}8~(1+r+\dots)$ and the fact that $x^2=rw\simeq{r^3}/8$ suggests that for large $N$ the optimal values of $r$ and $w(r)$ tend to zero.
Accounting for the choice~\eqref{e.quarticP} one can simplify the denominator of the distribution~\eqref{e.Pk} and write it as
\begin{equation}
\begin{aligned}
 \P_k &= \frac1{N{+}2\psi_{k}'}~\times
 \\
  &~~ \frac{2(2{-}r)rw}{(2{-}r)^2w^2+32(1{-}r)
			(w\tan^2\!{q}{+}2\sin^2\!q)\sin^2\!{q}} ~,
\end{aligned}
\label{e.Pk1}
\end{equation}
where $w=w(r)$ as given as in Eq.~\eqref{e.wr} and $k=2q$. This density can be expanded, taking into account Eq.~\eqref{e.wr}, and the relevant terms are
\begin{equation}
 \P_k \simeq \frac1N~\frac{4r^3/8}{r^4/16+64q^4}
 \simeq \frac{\sqrt{2}}N~ \frac{\Delta^3}{\Delta^4+q^4} ~,
\label{e.Pkquartic}
\end{equation}
where the half-width at half maximum is now defined as
\begin{equation}
 \Delta \equiv \frac{r}{4\sqrt{2}}~.
\end{equation}
Proceeding as in Sec.~\ref{ss.x.asymptotic}, the large-$N$ transition amplitude can then be transformed into the integral
\begin{equation}
 \alpha_\infty(t) = \frac{2\sqrt{2}\Delta^3}\pi \int\limits_0^{\pi/2} dq~
    \frac{\cos\{2[t(q{-}\sin{q}){-}sq{+}\psi_{2q}]\}}{\Delta^4+q^4}~,
\end{equation}
where again the arrival time is $t\,{=}\,N\,{+}\,s$, with the delay $s$. Note the qualitative difference represented by the quartic distribution.
Expanding Eqs.~\eqref{e.IkRk} gives $I_k\simeq{kr}$ and $R_k\simeq{2w{-}2k^2}$, so the asymptotic behavior of the phase shift~\eqref{e.psik} is
\begin{equation}
\psi_{2q} ~\simeq~ \tan^{-1}\!\frac{2qr}{r^2/4{-}8q^2}
~\simeq~ \tan^{-1}\!\frac{\sqrt{2}\,\xi}{1-\xi^2} ~,
\end{equation}
which replaces the second of Eqs.~\eqref{e.leading} and here again $q\,{=}\,\Delta\xi$.
Using the rescaled variables~\eqref{e.scaling} gives
\begin{equation}
 \alpha_\infty(t) = \frac{2\sqrt{2}}\pi\int\limits_0^\infty d\xi~
	\frac{\cos\big(\tau\xi^3{-}\sigma\xi+2\tan^{-1}\!\frac{\sqrt{2}\,\xi}
	{1-\xi^2}\big)} {1+\xi^4} ~.
\label{e.uinftyxi}
\end{equation}
It is again convenient, for the numerical evaluation, to perform the substitution $\xi=\tan{\zeta}$\,, so
\begin{equation}
 \frac{2\xi}{1{-}\xi^2} = \frac{2\tan\zeta}{1{-}\tan^2\!\zeta} = \tan{2\zeta}~,
\end{equation}
yielding
\begin{equation}
\begin{aligned}
 \alpha_\infty(\tau,\sigma) &= \frac{2\sqrt2}{\pi}\int\limits_0^{\pi/2}d\zeta~
		\frac{1{+}\tan^2\!\zeta}{1{+}\tan^4\!\zeta}\,\cos\theta(\zeta)
\\
  \theta(\zeta)&\equiv\tau\tan^3\!\zeta-\sigma\tan\zeta
		+2\tan^{-1}\!{\textstyle\frac{~\tan{2\zeta}}{\sqrt2}}~;
\end{aligned}
\label{e.uinftyz}
\end{equation}
the same integral appears, though in a different context, in Ref.~\onlinecite{ABCVV2012}.
\begin{figure}
\includegraphics[width=0.47\textwidth]{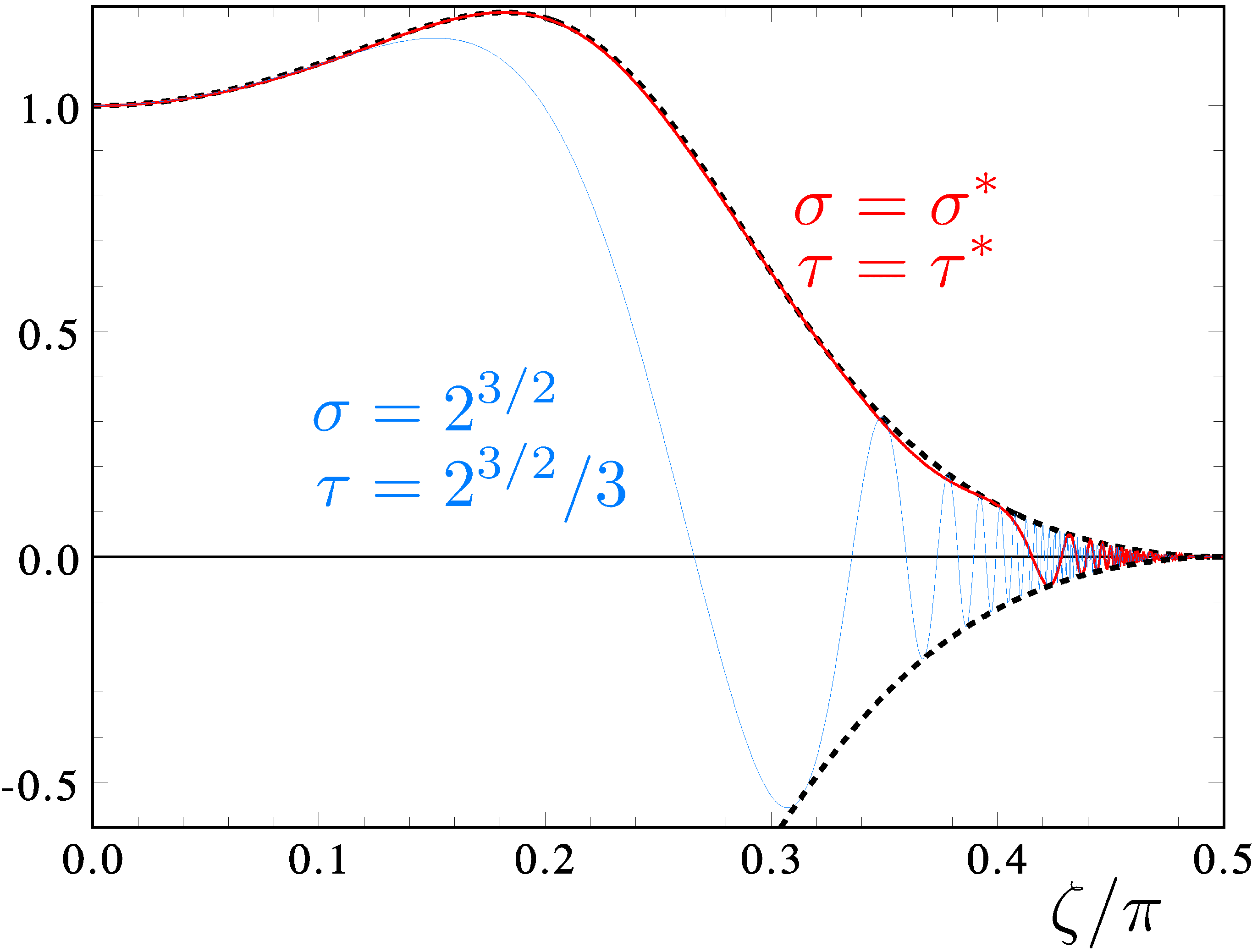}
\caption{Integrand of Eq.~\eqref{e.uinftyz} for the parameter pair $(\tau,\sigma)$ that cancels the linear and cubic terms in the phase of Eq.~\eqref{e.uinftyz} (blue, thin line) and that (red, thick line) which maximizes $\alpha_\infty(\sigma,\tau)$. The	dashed curve is the weighting function.}
\label{f.int_rw}
\end{figure}
At variance with Eq.~\eqref{e.ucos}, this sum of phase factors $\cos\theta(\zeta)$ is weighted by a prefactor that becomes small, $\sim(\tan\zeta)^{-2}$, in the region $\zeta\gtrsim\pi/4$ of large oscillations, so it is to be expected that larger values of $\alpha_\infty$ can be attained. In order to give an idea of the kind of integral one is dealing with, Fig.~\ref{f.int_rw} reports the integrand of Eq.~\eqref{e.uinftyz} for the values which eliminate first and third order terms of $\theta(\zeta)$, namely $\sigma=2^{3/2}$ and $\tau=2^{3/2}/3$, and for those values which yield the global maximum. The (algebraic) area below the curves is just $\alpha_\infty(\tau,\sigma)$. It is clear that the main point is in avoiding for as long as possible the onset of the rapid oscillations rather than setting the phase close to zero for small $\zeta$. It turns out that for any fixed $\sigma$, $\alpha_\infty(\tau,\sigma)$ has a maximum, i.e., the loss $\delta_\infty(\tau,\sigma)=1-\alpha_\infty(\tau,\sigma)$ reported in Fig.~\ref{f.minimiz} has a minimum at $\tau=\tau_{\rm{m}}(\sigma)$.
Repeating for different $\sigma$ one can find that the overall minimum of $\delta_\infty\big[\sigma,\tau_{\rm{m}}(\sigma)\big]$ is
\begin{equation}
 \delta_\infty^*=1-\alpha_\infty^{*}=0.012847~,
\end{equation}
corresponding to the asymptotic transmission amplitude
\begin{equation}
 \alpha_\infty^{*}=0.987153~,
\end{equation}
and occurs when
\begin{equation}
 \sigma^{*}\,{=}\,3.1645
~,~~~~
 \tau^{*}\,{\equiv}\,\tau_{\rm{m}}(\sigma^*)\,{=}\,0.15545~;
\end{equation}
as expected, using two adjustable parameters in the place of one strongly improved the transmission  found in Sec.~\ref{ss.x.asymptotic}. The scaling of Eq.~\eqref{e.scaling} tells that asymptotically
\begin{equation}
 \Delta^* \simeq 0.77548~N^{-1/3}
~,~~~~
 s^* \simeq 2.0403~N^{1/3}~.
\end{equation}
The optimal values of the parameters $r$ and $w$ behave as
\begin{equation}
 r^* \simeq~ 4.3868~N^{-1/3}
~,~~~~
 w^* \simeq~ 2.4055~N^{-2/3} ~,
\label{e.rw_N}
\end{equation}
entailing that
\begin{equation}
 x^* \simeq~ 3.2484~N^{-1/2}
~,~~~~
 y^* \simeq~ \sqrt{2}-1.5510~N^{-1/3} ~.
\label{e.xy_N}
\end{equation}
For the optimal masses and spring, Eq.~\eqref{e.mKxyz}, one finds
\begin{equation}
\begin{aligned}
 m_1^*    &= \frac{r}{(2{-}r)w} ~\simeq~  0.9118~N^{1/3}
\\
 m_2^*    &= \frac1{2{-}r}      ~\simeq~ \frac12+1.0967~N^{-1/3}
\\
 K_{12}^* &= \frac{r}{2{-}r}    ~\simeq~  2.1934~N^{-1/3}~.
\end{aligned}
\label{e.m1m2K12-rw}
\end{equation}
Therefore, the best transmitting chain has a larger first mass, while
the second is about one half the bulk ones, and the spring connecting them
has to be weak, inversely proportional to $m_1$.
The above asymptotic behaviors are reported in Figs.~\ref{f.delta}, \ref{f.x_xy} and~\ref{f.m_mmK}.

Note that without the assumption~\eqref{e.quarticP} the density denominator would have a quadratic term and the dominant part, Eq.~\eqref{e.Pkquartic}, would have the same form of the preceding section, leading to a similar asymptotic integral and to a much worse limit of the transmission amplitude: This justifies the elimination of the $k^2$-terms by means of the constraint~\eqref{e.wr}, as also confirmed by the numerical results presented below.

\subsection{Numerical results for finite $N$}

\begin{figure}
\includegraphics[width=0.47\textwidth]{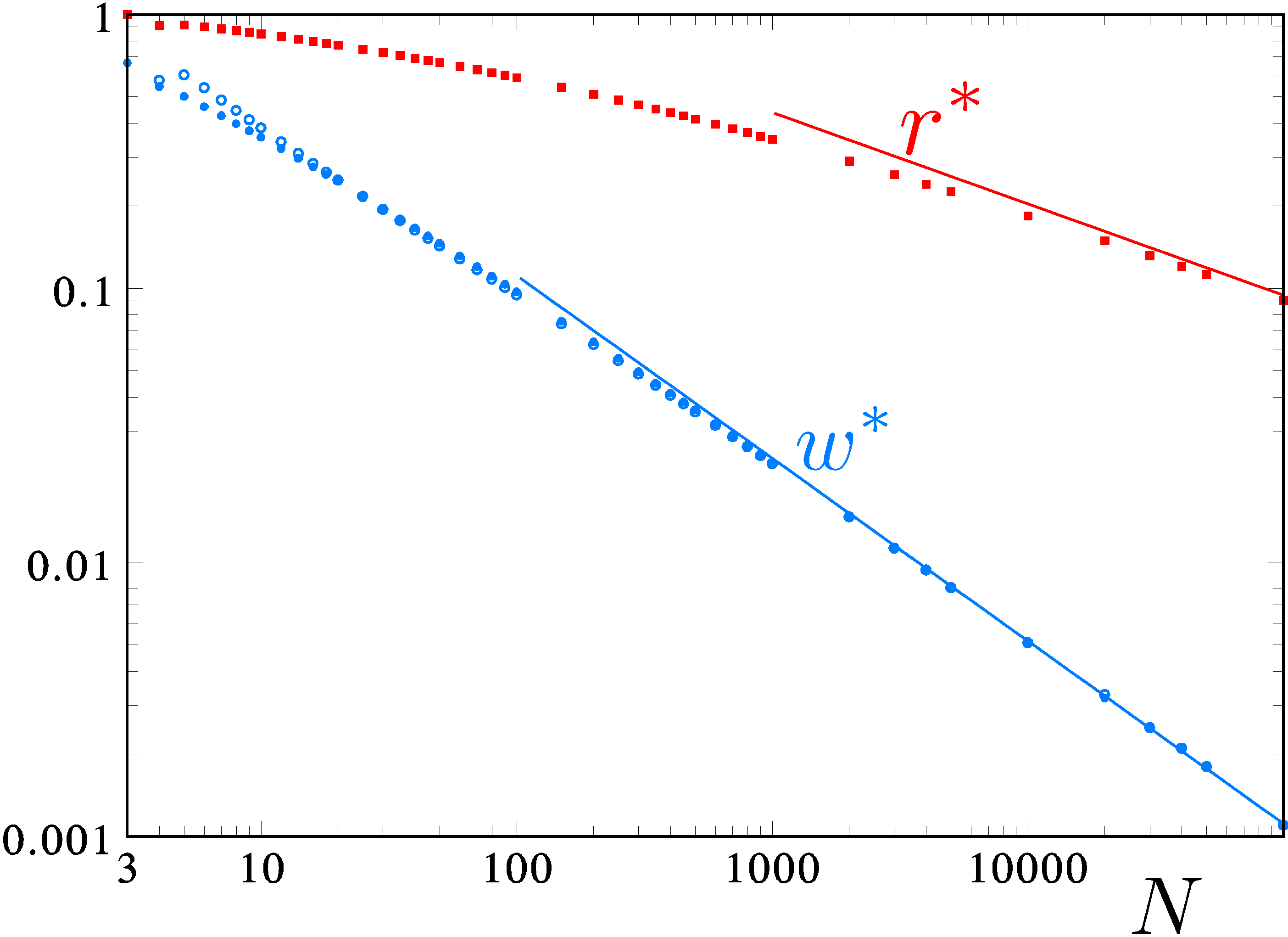}
\caption{Optimal parameters $r^*$ (red squares) and $w^*$ (blue circles) reported vs. $N$ for the chain of Sec.~\ref{s.m1m2K12}. The curves are the large-$N$ asymptotics of Eq.~\eqref{e.rw_N}. The open circles are the values of the function $w(r^*)$, Eq.~\eqref{e.wr}: For $N\gtrsim20$ they are almost indistinguishable from the optimal values $w^*$ confirming that optimal transmission requires a quartic mode density.}
\label{f.rw}
\end{figure}

The optimal values of $r$ and $w$ which maximize the transmission amplitude have been numerically evaluated for different values of $N$, as reported in Table~\ref{t.opt_rw}. Note that for each pair $(r,w)$ the algorithm has first to evaluate the allowed $k_n$, which is done for each $n$ by iterating Eq.~\eqref{e.kn1} starting from the estimate $k_n=\pi(n{-}1)/N$ and using the expression~\eqref{e.psik} for the shift $\psi_k$; then, the maximum of $\alpha_{_N}(t)$, Eq.~\eqref{e.utn}, is found by scanning on the arrival delay $s=t{-}N$ and taking the maximal value $\alpha_{_N}(N{+}s)$.
By studying these values on the $(r,w)$ plane (i.e., the numerical code contains three nested loops for $s$, $w$, and $r$, respectively) the optimized transmission maximum $\alpha^*(N{+}s^{*})$ corresponding to the values  $r^{*}$ and $w^{*}$ has been found. The numerical outcomes shown in Fig~\ref{f.rw} confirm the conjecture that $w^*$ approaches, for large $N$, the value $w(r^*)$ which makes the mode density quartic. The corresponding transmission loss is reported in Fig.~\ref{f.delta}: At variance with the monotonic increase with $N$ found in the preceding section, the transmission loss displays maximum $\delta^*\simeq0.0157$ at about $N\simeq{60}$, followed by a decrease towards the asymptotic value. The related parameters $x^*$ and $y^*$, as well as those of the two masses and spring are shown in Figs.~\ref{f.x_xy} and~\ref{f.m_mmK}, respectively. 

\begin{figure}
\includegraphics[width=0.47\textwidth]{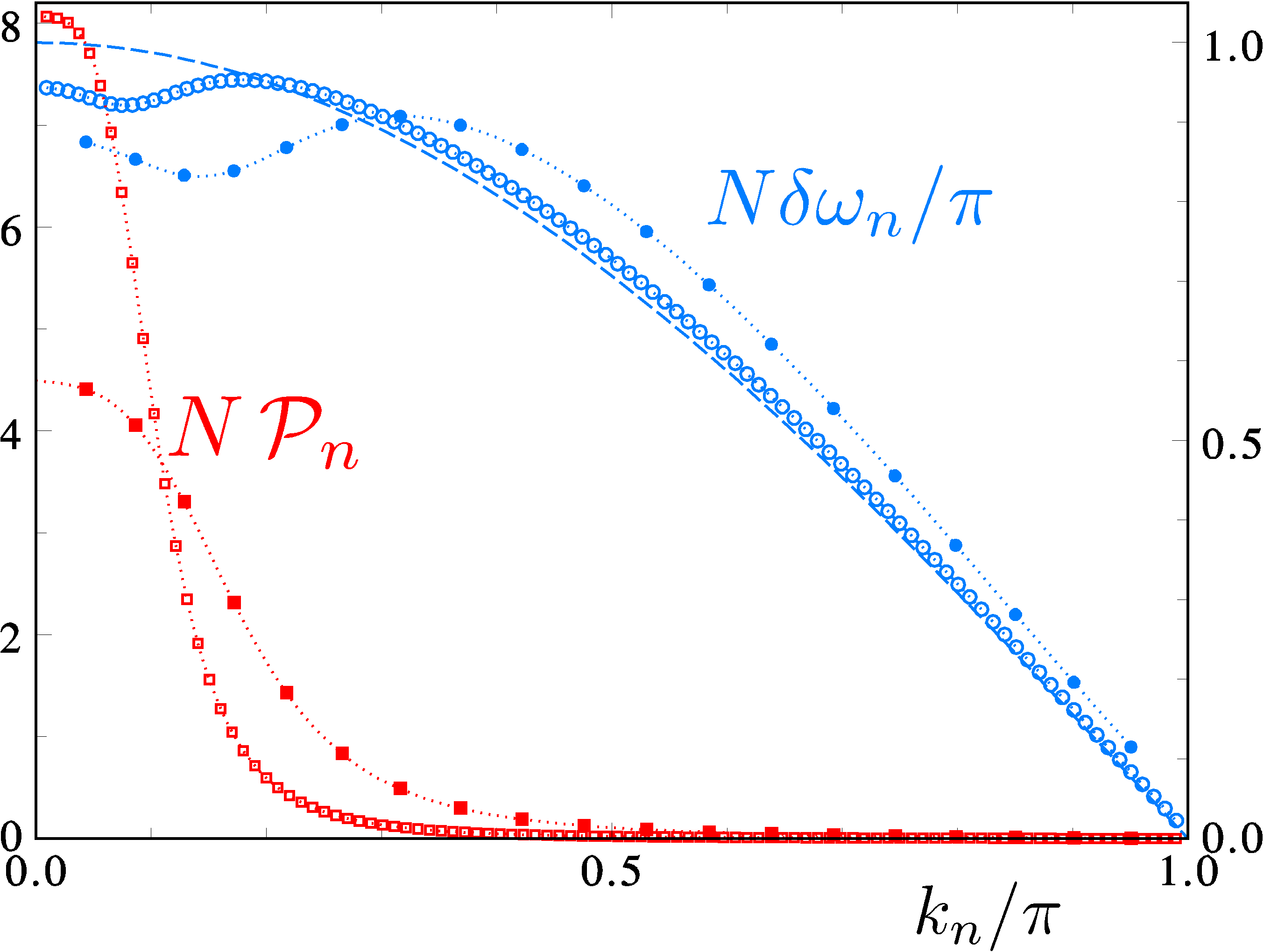}
\caption{Mode density $\P_n$ (red squares, left $y$ axis) and frequency spacing $\delta\omega_n=\omega_n\,{-}\,\omega_{n-1}$ (blue circles, right $y$ axis) for $N=20$ (filled symbols) and $N=100$ (open symbols) in the optimized case with parameters $(r^*,w^*)$ as in Table~\ref{t.opt_rw}. The dashed curve is the uniform-chain group velocity $v_k=\cos\frac{k}2$. Since the mode density has tiny tails, only modes with almost equally spaced frequencies are involved in the dynamics.}
\label{f.Pk20-100}
\end{figure}

In Fig.~\ref{f.Pk20-100} the shapes of the optimal mode density for $N=20$ and $N=100$ can be compared with the exact frequency spacings $\delta\omega_n=\omega_n\,{-}\,\omega_{n-1}$, explaining the basic mechanism of coherence: Lowering $r$ shrinks the density involving smaller-$k$ modes, but also deforms the frequency spacings, and there is a best compromise between the two effects. Note that $N\delta\omega_n/\pi$ corresponds to the ``group velocity''~\eqref{e.vk}, whose value for the uniform chain is reported as a dashed line.

\section{Conclusions}
\label{s.concl}
Highly efficient pulse transmission over a uniform discrete elastic system, modelized as a chain of $N$ masses connected by springs, has been shown to be possible irrespectively of $N$ just by symmetrically modifying, at both chain ends, two masses and the spring between them. This result is far from trivial, since the dynamics of such a chain is almost invariably affected by strong dispersion. The main point is in the analysis of the normal modes which are excited by the initial pulse: Their frequencies are made almost equally spaced, yielding coherent transmission, by suitably tuning two free parameters.

A pulse starting from an end of such an elastic chain can bounce back and forth several times before being ``absorbed'' by dispersion. This suggests the possibility of creating a mechanical toy similar but alternative to the Newton cradle, e.g., like that depicted in Fig.~\ref{f.cradle4}. Its behavior would be somewhat bizarre due to the large duration of the bounces, of the order of $N\sqrt{m/K}$ (where $m$ and $K$ are the bulk masses and elastic constants), as it would work also for a long spring-mass chain.

\begin{figure}
\includegraphics[width=0.47\textwidth]{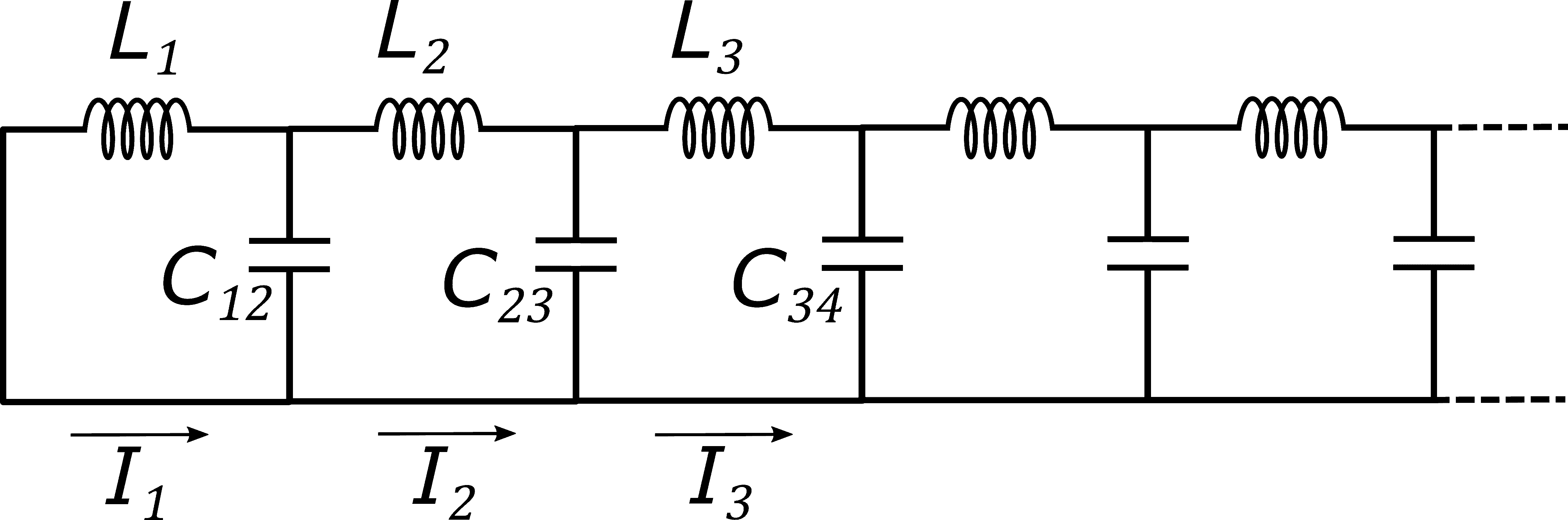}
\caption{The electric circuit equivalent to the spring-mass chain~\eqref{e.Horig}: Capacitances replace the elastic constants, $C_{i,i+1}\leftrightarrow{K^{-1}_{i,i+1}}$, and inductances replace the masses $L_i\leftrightarrow{m_i}$ .}
	\label{f.LCcircuit}
\end{figure}

The array of electric LC circuits depicted in Fig.~\ref{f.LCcircuit} can also be described by the Hamiltonian~\eqref{e.Horig}, with capacitors and inductors replacing springs and masses, respectively, i.e., setting
\begin{equation}
 K_{i,i+1} = C_{i,i+1}^{-1}
~,~~~~~~
 m_i = L_i ~;
\end{equation}
the role of the coordinate is played by the charge $Q_i$ flowing in each loop yielding the current $I_i=\dot{Q}_i$. For instance, a current pulse given, e.g., by an external inductive coupling to $L_1$, would travel back and forth through the electric array, provided the extremal inductances and capacitances are suitably tuned, while having the bulk uniform values $L$ and $C$. The time scale is $\sqrt{LC}$ and the end-to-end time travel is about $N$ times larger.

Other, more useful, applications can exploit the capability of transferring a concentrated amount of energy between the chain ends; for instance, one can imagine a wall, constituted of an array of elastically coupled layers, that could efficiently transmit heath or sound. The same idea can be transferred to the nanoscale~\cite{CahillEtAl2003,NorrisLB2013}, for instance, to atomic chains~\cite{RammPH2014} or multilayers. For instance, ballistic phonon transport could be engineered leading to different behavior, e.g., of the thermal conductance. Finally, it is a suggestive idea that the same simple mechanism be able to explain some features of energy transport in biological structures.

\acknowledgements
The author gratefully thanks L.~Banchi, T.~J.~G.~Apollaro, A.~Cuccoli, and P.~Verrucchi for fruitful discussions.

\appendix

\section{Perfect $5$-mass chain}
\label{a.N5}

For $N=5$ the interaction matrix~\eqref{e.BN} reads
\begin{equation}
\bm{B} =
\begin{bmatrix}
	~w~ & -x   &  0  &  0  & 0  \\
	-x  &  t   & -y  &  0  & 0  \\
	 0  & -y   &  2  & -y  & 0  \\
	 0  &  0   & -y  &  t  & -x \\
	 0  &  0   &  0  & -x  &  w  
\end{bmatrix}~~,
\label{e.N4}
\end{equation}
where
\begin{equation}
\begin{aligned}
 &w=\frac{K_{12}}{m_1} ~,
 &&t=\frac{1+K_{12}}{m_2} ~,
\\
 &x=\frac{K_{12}}{\sqrt{m_1m_2}} ~,
 &&y=\frac1{\sqrt{m_2}} ~,
\end{aligned}
\end{equation}
and one actually has three free parameters, since $w(t{-}y^2)={x^2}$. This is more general than the model~\eqref{e.xyz}, which assumes $t=2$. To proceed, it is convenient to set $r\equiv{t}{-}y^2$, so $x^2=wr$. Calculated by expanding in the middle row or column, the determinant giving the secular equation is obtained in factorized form,
\begin{equation}
 0=\det(\mu\,{-}\,\bm{B})=P(\mu)\,Q(\mu)
\end{equation}
with 
\begin{equation}
\begin{aligned}
 P(\mu) &= (\mu{-}w)(\mu{-}t)-x^2
\\
 &= \mu^2-(w{+}t)\mu+wy^2
\\
 &\equiv (\mu-\mu_2)(\mu-\mu_4)
\\Q(\mu) &= (\mu{-}2)\big[[(\mu{-}w)(\mu{-}t)-x^2\big]
           -2(\mu{-}w)y^2
\\
  &= \mu\big[ \mu^2-(w{+}t{+}2)\mu+wy^2+2(w{+}r)\big]
\\
  &\equiv \mu(\mu-\mu_3)(\mu-\mu_5)
~;
\end{aligned}
\end{equation}
the eigenvalues $\{\mu_n,~n=1,\,...,\,5\}$ are in increasing order and to obtain equally spaced frequencies $\omega_n=\sqrt{\mu_n}$ one imposes $\mu_n=(n{-}1)^2\,u$, which entails
\begin{equation}
\left\{
\begin{aligned}
 \mu_2+\mu_4&=10u=w+t
\\
 \mu_3+\mu_5&=20u=w+t+2
\end{aligned}
\right.
~\Longrightarrow~
\left\{
\begin{aligned}
 &u={\textstyle \frac15}
\\
 &w+t=2
\end{aligned}
\right.~.
\end{equation}
and also
\begin{equation}
\left\{
\begin{aligned}
 \mu_2\mu_4&=9u^2=wy^2
\\
 \mu_3\mu_5&=64u^2=wy^2{+}\,2(w{+}r)
\end{aligned}
\right.
~\Longrightarrow~
\left\{
\begin{aligned}
 &wy^2={\textstyle \frac9{25}}
\\
 &w+r={\textstyle \frac{11}{10}}
\end{aligned}
\right.~.
\end{equation}
It is then straightforward to obtain
\begin{equation}
 w={\textstyle\frac25} ~,~~
 t={\textstyle\frac85} ~,~~
 r={\textstyle\frac7{10}} ~,~~
 x^2={\textstyle\frac7{25}} ~,~~
 y^2={\textstyle\frac9{10}} ~,
\end{equation}
and the result~\eqref{e.N5} follows.

\section{Mode density}
\label{a.modedensity}

The mode density~\eqref{e.Pn} is given by the square components of the first column of the orthogonal diagonalizing matrix $U_{ki}$. For this quantity Parlett reports a very useful formula (Corollary 7.9.1 of Ref.~\onlinecite{Parlett1998}), namely
\begin{equation}
\P_k \equiv U_{k1}^2
= \frac{\chi_{_{2:N}}(\lambda)}{\partial_\lambda\chi_{_N}(\lambda)}
= -\frac{2\sin{k}~\chi_{_{2:N}}(k)}{\partial_k\chi_{_N}(k)}~,
\label{e.Uk1}
\end{equation}
where $\chi_{_{2:N}}(k)$ is the first minor of the determinant~\eqref{e.chiN} and $k$ is a root of the characteristic polynomial, $\chi_{_N}(k)=0$. The minor coincides with the part that multiplies $(\lambda{-}z)$ in Eq.~\eqref{e.chixi},
\begin{equation}
\begin{aligned}
	\chi_{_{2:N}} &= \lambda\xi_{_{N-2}} - y^2\xi_{_{N-3}}
	\\
	&=  \lambda\big[(\lambda^2{-}z\lambda{-}x^2)\eta_{_{N-4}}
	-(\lambda{-}z)y^2\eta_{_{N-5}}\big]-
	\\
	& ~~~~~ -y^2\big[(\lambda^2{-}z\lambda{-}x^2)\eta_{_{N-5}}
	-(\lambda{-}z)y^2\eta_{_{N-6}}\big] ~;
\end{aligned}
\end{equation}
in the second line Eqs.~\eqref{e.xieta} has been used, and then from Eq.~\eqref{e.chebyshev},
\begin{equation}
\begin{aligned}
	\sin{k}~\chi_{_{2:N}} &= 
	\Im\big\{e^{i(N{-}3)k} (\lambda-y^2e^{-ik})~\times
	\\
	& ~~~~~\times \big[
	(\lambda^2{-}z\lambda{-}x^2)-(\lambda{-}z)y^2e^{-ik}\big]\big\}
	\\
	&= \Im\big\{e^{iNk} a_ku_k\big\}~,
\end{aligned}
\end{equation}
where
\begin{equation}
a_k \equiv e^{-ik}(\lambda-y^2e^{-ik}) ~.
\label{e.ak}
\end{equation}

By deriving Eq.~\eqref{e.chik} with respect to $k$ one has
\begin{equation}
\begin{aligned}
	\partial_k &\big[\chi_{_N}(k)\sin{k}\big]
	= \sin{k}~\partial_k\chi_{_N}(k)
	\\
	&= \Im\big\{e^{i(N{+}1)k}~[i(N{+}1)u_k^2+2u_ku'_k]\big\}
	\\
	&= (N{+}1)\,\Re\big\{e^{i(N{+}1)k}u_k^2 \big\}
	+2\,\Im\big\{e^{i(N{+}1)k}u_ku'_k\big\};
\end{aligned}
\end{equation}
the secular equation entails that the argument of $\Re$ is real and
$\Re$ can be omitted; then,
\begin{equation}
\begin{aligned}
	\sin&{k}~\partial_k\chi_{_N}(k)
	= \Big[N{+}1+2\,\Im\Big\{
	\frac{u'_k}{u_k}\Big\}\Big]\,e^{i(N{+}1)k}\,u_k^2
	\\
	&= \Big[N{+}1+2\,\Im\big\{
	\partial_k\ln{u_k}\big\}\Big]\,e^{i(N{+}1)k}\,u_k^2
	\\
	&= \big[N{+}1 +2\,\Im\big\{
	\partial_k\ln|u_k| -i\varphi_k'\big\}\big]\,e^{i(N{+}1)k}\,u_k^2
	\\
	&= (N{+}1-2\varphi_k')\,e^{i(N{+}1)k}\,u_k^2~.
\end{aligned}
\label{e.dkchi}
\end{equation}
The Parlett formula~\eqref{e.Uk1} becomes
\begin{equation}
\begin{aligned}
	\P_k &= -\frac{2\sin{k}}{N{+}1{-}2\varphi_k'}
	~\frac{\Im\{e^{iNk}a_ku_k\}}{e^{i(N{+}1)k}u_k^2}
	\\
	&= \frac{2\sin{k}}{N{+}1{-}2\varphi_k'}
	~ \frac{\Im\{e^{ik}u_ka_k^*\}}{|u_k|^2} ~;
\end{aligned}
\label{e.Uk1a}
\end{equation}
comparing Eqs.~\eqref{e.uk1} and~\eqref{e.ak} one has 
\begin{equation}
u_ke^{2ik}=(\lambda-z)a_ke^{ik}-x^2
\end{equation}
and it follows
\begin{equation}
\begin{aligned}
	\Im\{e^{ik}u_ka_k^*\} &= \Im\{(u_ke^{2ik})(a_ke^{ik})^*\}
	\\
	&= \Im\{(\lambda-z)|a_k|^2-x^2(\lambda-y^2e^{ik})\}
	\\
	&= x^2y^2\,\sin{k} ~.
\end{aligned}
\end{equation}
This yields Eq.~\eqref{e.Uk1b}.  Note that
\begin{equation}
\begin{aligned}
	u_k^*\,u_k'
	&= u_k^*\,\partial_k \big(|u_k|\,e^{-i\varphi_k}\big)
	\\
	&= u_k^*\,e^{-i\varphi_k}\big(\partial_k |u_k|-i |u_k| \varphi_k'\big)
	\\
	&= |u_k|\partial_k |u_k|-i |u_k|^2 \varphi_k'
\end{aligned}
\end{equation}
implies the identity
\begin{equation}
|u_k|^2 \varphi_k' = - \Im\{u_k^*\,u_k'\} ~,
\label{e.phiprimed}
\end{equation}
so the density can also be written in the alternative form
\begin{equation}
 \P_k = \frac{2x^2y^2\sin^2\!{k}}{(N{+}1)|u_k|^2{+}2\Im\{u_k^*\,u_k'\}} ~.
\end{equation}

\section{Explicit expression of $\P_k$}
\label{a.density_rw}

From Eq.~\eqref{e.tudef} one writes the real and imaginary parts of $\tu_k\equiv{R_k+iI_k}$,
\begin{equation}
\begin{aligned}
	R_k &\equiv 1{-}(1{-}w)\cos{k}{-}(1{-}w)(1{-}r)\cos{2k}{+}(1{-}r)\cos{3k}
	\\
	I_k &\equiv (1{-}w)\sin{k}+(1{-}w)(1{-}r)\sin{2k}-(1{-}r)\sin{3k}
\end{aligned}
\label{e.IkRk}
\end{equation}
and since~\cite{note-atan2}
\begin{equation}
 \psi_k=\tan^{-1}\frac{I_k}{R_k}
\label{e.psik}
\end{equation}
it follows that
\begin{equation}
|\tu_k|^2\psi_k' = R_kI_k'-R_k'I_k ~.
\end{equation}
An expression analogous to Eq.~\eqref{e.phiprimed} holds for
$\psi_k'$,
\begin{equation}
\begin{aligned}
	\tu_k^*\,\tu_k'
	&= \tu_k^*\,\partial_k \big(|\tu_k|\,e^{i\psi_k}\big)
	\\
	&= \tu_k^*\,e^{i\psi_k}\big(\partial_k |\tu_k|+i |\tu_k| \psi_k'\big)
	\\
	&= |\tu_k|\partial_k |\tu_k|+i |\tu_k|^2 \psi_k' ~,
\end{aligned}
\end{equation}
while Eq.~\eqref{e.ukz0} entails $|u_k|^2=4\sin^2\!\fr{k}2\,|\tu_k|^2$; eventually, the density~\eqref{e.Uk1b} turns into
\begin{equation}
\P_k = \frac{2(2{-}r)rw}{N{+}2\psi_k'}~\frac{\cos^2\!\fr{k}2}{|\tu_k|^2}
= \frac{2(2{-}r)rw\,\cos^2\!\fr{k}2}
{N|\tu_k|^2{+}2\,\Im\{\tu_k^*\,\tu_k'\}} ~.
\label{e.Uk1d}
\end{equation}
Setting $k\equiv2\,q$ and using the identities
\begin{equation}
\begin{aligned}
	\cos{3q} &=  \cos{q}~(1-4\sin^2\!q) ~,
	\\
	\sin{3q} &=  \sin{q}~(4\cos^2\!q-1) ~,
\end{aligned}
\end{equation}
one has
\onecolumngrid
\phantom{.}
\begin{equation}
\begin{aligned}
	e^{3iq}\tu_k
	&= \big[e^{3iq}+(1{-}r)e^{-3iq}\big]-(1{-}w)\big[e^{iq}+(1{-}r)e^{-iq}\big]
	\\
	&= \big[(2{-}r)\cos{3q}+ir\sin{3q}\big]-(1{-}w)\big[(2{-}r)\cos{q}+ir\sin{q}\big]
	\\
	&= (2{-}r)\big(w-4\sin^2\!{q}\big)\cos{q}-ir\big(2{-}w-4\cos^2\!{q}\big)\sin{q} ~;
	\label{e.tukz0}
\end{aligned}
\end{equation}
from this one can work out the square modulus needed to calculate Eq.~\eqref{e.Uk1d},
\begin{equation}
\begin{aligned}
	|\tu_k|^2
	&= (2{-}r)^2\big(w-4\sin^2\!{q}\big)^2\cos^2\!{q}
	+r^2\big(2{-}w-4\cos^2\!{q}\big)^2\sin^2\!{q}
\\
	&= (2{-}r)^2\big[w^2-8(w{-}2\sin^2\!{q})\sin^2\!{q}\big]\cos^2\!{q}
	+r^2\big[(2{-}w)^2-8(2{-}w{-}2\cos^2\!{q})\cos^2\!{q}\big]\sin^2\!{q}
\\
	&= \big[(2{-}r)^2w^2+r^2(2{-}w)^2\tan^2\!{q}
	-32(1{-}r)(w{-}2\sin^2\!{q})\sin^2\!{q}\big]\cos^2\!{q}
	\label{e.tu2}
\end{aligned}
\end{equation}
Eventually, from Eq.~\eqref{e.Uk1d},
\begin{equation}
\P_k = \frac1{1+\delta_{k0}}\,\frac1{N{+}2\psi_{k}'}~
\frac{2(2{-}r)rw}{(2{-}r)^2w^2+r^2(2{-}w)^2\tan^2\!q
	-32(1{-}r)(w{-}2\sin^2\!q)\sin^2\!q}  ~,~~~~ \Big(q\equiv\frac k2\,\Big) ~.
\label{e.Pk}
\end{equation}
\twocolumngrid
The first term halves the value of $\P_0$ and is needed to restore the correct distribution. The reason for it can be more easily understood in the uniform limit $w\,{=}\,r\,{=}\,1$, where one obtains the density~\eqref{e.Pkunif}, but with the correct value for $k\,{=}\,0$ being $\P_0\,{=}\,\frac1N$. Indeed, an accurate derivation of the free-chain limit $z\,{\to}\,1^-$ gives for the lowest $k$-mode the shift $\psi_{k_1}\,{\simeq}\,-(1{-}z)/k_1$, and by Eq.~\eqref{e.kn1} one has $k_1\,{\simeq}\,\sqrt{2(1{-}z)/N}\to{0}$; therefore, all shifts and their derivatives $\psi_{k}'$ tend to zero, but with the exception of a finite limit for $2\psi'_{k_1}\,{\simeq}\,2(1{-}z)/k_1^2\,{\to}\,N$. Actually, one can analytically verify that with this caveat the distribution~\eqref{e.Pkunif} is correctly normalized. It has been numerically checked that the more general distribution~\eqref{e.Pk} requires the the value at $k{=}0$ to be halved in order to be exactly normalized.


\onecolumngrid

\begin{table}[t]
\caption{Quasiuniform chain: Numerical optimization of the only extremal masses $m_1=m_N$, considered in Sec.~\ref{s.m1}. For different $N$ the reported quantities are the optimal parameter $x^{*}$ that minimizes  the transmission loss $\delta^{*}(x)\equiv\,1{-}\alpha^*(x)$, the corresponding time delay $s^{*}\equiv{t}^{*}\,{-}\,N$, mass $m_1^{*}$, and width parameter $\Delta^{*}$. For comparison, the last column reports the transmission loss for the fully uniform chain, i.e, for $x\,{=}\,1$. Note that for $N=3$ the outcome agrees with the perfect transmission analytically found in Sec.~\ref{ss.N3N4}.}
\label{t.opt_y1}
\medskip
\begin{tabular}{|r@{\hskip 2mm}|ccr@{\hskip 2mm}|c@{\hskip 2mm}l|c|}
\hline
   $N$ & $x^{*}$& $\delta^*(x^*)$ & $s^{*}(x^*)$ & $m_1^{*}$ & ~~$\Delta^{*}$ & $~~~\delta^*(1)$~~~ \\
		\hline
     3 & 0.8165 & 0.0000 &  0.85 &  1.500 & 0.5000  & 0.0292 \\
     4 & 0.7731 & 0.0041 &  1.08 &  1.673 & 0.4261  & 0.0608 \\
     5 & 0.7414 & 0.0083 &  1.28 &  1.819 & 0.3790  & 0.0900 \\
     6 & 0.7171 & 0.0123 &  1.44 &  1.945 & 0.3461  & 0.1161 \\
     7 & 0.6975 & 0.0160 &  1.58 &  2.056 & 0.3214  & 0.1395 \\
     8 & 0.6811 & 0.0193 &  1.71 &  2.156 & 0.3020  & 0.1606 \\
     9 & 0.6670 & 0.0223 &  1.83 &  2.248 & 0.2861  & 0.1796 \\
    10 & 0.6548 & 0.0251 &  1.94 &  2.333 & 0.2728  & 0.1969 \\
    12 & 0.6341 & 0.0299 &  2.13 &  2.487 & 0.2517  & 0.2274 \\
    14 & 0.6173 & 0.0341 &  2.31 &  2.625 & 0.2353  & 0.2535 \\
    16 & 0.6031 & 0.0377 &  2.46 &  2.750 & 0.2223  & 0.2762 \\
    18 & 0.5908 & 0.0409 &  2.61 &  2.865 & 0.2114  & 0.2962 \\
    20 & 0.5801 & 0.0437 &  2.74 &  2.972 & 0.2023  & 0.3141 \\
    25 & 0.5580 & 0.0496 &  3.04 &  3.212 & 0.1844  & 0.3516 \\
    30 & 0.5406 & 0.0544 &  3.30 &  3.421 & 0.1711  & 0.3817 \\
    35 & 0.5263 & 0.0583 &  3.54 &  3.610 & 0.1608  & 0.4067 \\
    40 & 0.5143 & 0.0617 &  3.75 &  3.781 & 0.1524  & 0.4280 \\
    45 & 0.5039 & 0.0646 &  3.94 &  3.938 & 0.1454  & 0.4464 \\
    50 & 0.4948 & 0.0671 &  4.12 &  4.085 & 0.1395  & 0.4626 \\
    60 & 0.4794 & 0.0714 &  4.45 &  4.352 & 0.1298  & 0.4898 \\
    70 & 0.4667 & 0.0749 &  4.75 &  4.590 & 0.1222  & 0.5122 \\
    80 & 0.4561 & 0.0778 &  5.01 &  4.808 & 0.1161  & 0.5310 \\
    90 & 0.4469 & 0.0804 &  5.26 &  5.008 & 0.1109  & 0.5471 \\
   100 & 0.4388 & 0.0826 &  5.49 &  5.194 & 0.1065  & 0.5612 \\
   150 & 0.4090 & 0.0906 &  6.45 &  5.977 & 0.09129 & 0.6121 \\
   200 & 0.3892 & 0.0958 &  7.21 &  6.602 & 0.08195 & 0.6452 \\
   250 & 0.3745 & 0.0996 &  7.86 &  7.130 & 0.07541 & 0.6692 \\
   300 & 0.3629 & 0.1026 &  8.42 &  7.593 & 0.07049 & 0.6877 \\
   350 & 0.3534 & 0.1049 &  8.93 &  8.008 & 0.06660 & 0.7027 \\
   400 & 0.3454 & 0.1069 &  9.39 &  8.385 & 0.06342 & 0.7151 \\
   450 & 0.3384 & 0.1086 &  9.81 &  8.731 & 0.06074 & 0.7256 \\
   500 & 0.3323 & 0.1100 & 10.20 &  9.054 & 0.05845 & 0.7348 \\
   600 & 0.3221 & 0.1124 & 10.92 &  9.639 & 0.05471 & 0.7499 \\
   700 & 0.3137 & 0.1144 & 11.56 & 10.164 & 0.05174 & 0.7621 \\
   800 & 0.3066 & 0.1160 & 12.14 & 10.640 & 0.04931 & 0.7722 \\
   900 & 0.3004 & 0.1174 & 12.67 & 11.079 & 0.04727 & 0.7807 \\
  1000 & 0.2951 & 0.1186 & 13.17 & 11.486 & 0.04551 & 0.7881 \\
  2000 & 0.2621 & 0.1255 & 16.91 & 14.559 & 0.03556 & 0.8311 \\
  3000 & 0.2446 & 0.1289 & 19.53 & 16.717 & 0.03083 & 0.8522 \\
  4000 & 0.2329 & 0.1310 & 21.63 & 18.436 & 0.02788 & 0.8656 \\
  5000 & 0.2242 & 0.1326 & 23.39 & 19.889 & 0.02579 & 0.8752 \\
 10000 & 0.1994 & 0.1368 & 29.80 & 25.158 & 0.02028 & 0.9008 \\
 20000 & 0.1773 & 0.1401 & 37.88 & 31.804 & 0.01597 & 0.9212 \\
 30000 & 0.1656 & 0.1417 & 43.55 & 36.465 & 0.01390 & 0.9311 \\
 40000 & 0.1578 & 0.1428 & 48.06 & 40.180 & 0.01260 & 0.9374 \\
 50000 & 0.1520 & 0.1435 & 51.87 & 43.311 & 0.01168 & 0.9419 \\
100000 & 0.1350 & 0.1455 & 65.76 & 54.870 & 0.00920 & 0.9539 \\
500000 & 0.1032 & 0.1486 & 113.3 & 93.895 & 0.00535 & 0.9730 \\
\hline
$N\to\infty$ & 0.9173\,$N^{-\frac16}$~ &  0.1531 & 1.444\,$N^{\frac13}$
        & 1.188\,$N^{\frac13}$ & 0.4207\,$N^{-\frac13}$ & 1.0000 \\
		\hline
	\end{tabular}
\end{table}

\setlength\tabcolsep{1mm}
\begin{table}[t]
\caption{Quasiuniform chain: Numerical optimization of the two extremal masses, $m_1=m_N$ and $m_2=m_{N-1}$, and of the spring in between $K_{12}=K_{N-1,N}$, considered in Sec.~\ref{s.m1m2K12}. For different $N$ the reported quantities are the optimal parameters $r^{*}$ and $w^{*}$ which minimize the transmission loss $\delta^{*}(r,w)\equiv\,1{-}\alpha^*(r,w)$, and the associated time delay $s^{*}\equiv{t}^{*}\,{-}\,N$. The last three columns report the corresponding optimal masses and elastic constant, Eq.~\eqref{e.m1m2K12-rw}. Note that for $N=3$ and $N=4$ perfect transmission is found, in agreement with Sec.~\ref{ss.N3N4}.}
\label{t.opt_rw}
\medskip
\begin{tabular}{|r@{\hspace{3mm}}|@{\hspace{3mm}}lllr@{\hspace{3mm}}|
	                 S[table-format=-3.4]@{\hspace{4mm}}l@{\hspace{3mm}}l|}
		\hline
		$N$ & $r^{*}$ & $w^{*}$ & $\delta^{*}(r^{*},w^{*})$ & $s^{*}$~~~ & \mbox{$m_1^{*}$} & $m_2^{*}$ & $K_{12}^{*}$ \\
		\hline
     3  &  1.0000  &  0.6667  &  0.00000  &   0.85  &  1.500  &  1.0000  &  1.0000 \\
     4  &  0.9091  &  0.5455  &  0.00000  &   1.21  &  1.528  &  0.9167  &  0.8333 \\
     5  &  0.9151  &  0.5024  &  0.00304  &   1.40  &  1.679  &  0.9218  &  0.8435 \\
     6  &  0.9003  &  0.4610  &  0.00505  &   1.59  &  1.776  &  0.9093  &  0.8187 \\
     7  &  0.8853  &  0.4273  &  0.00664  &   1.77  &  1.859  &  0.8971  &  0.7942 \\
     8  &  0.8718  &  0.3996  &  0.00794  &   1.93  &  1.933  &  0.8863  &  0.7727 \\
     9  &  0.8595  &  0.3764  &  0.00902  &   2.08  &  2.002  &  0.8768  &  0.7536 \\
    10  &  0.8483  &  0.3565  &  0.00991  &   2.22  &  2.066  &  0.8683  &  0.7366 \\
    12  &  0.8285  &  0.3241  &  0.01130  &   2.47  &  2.182  &  0.8536  &  0.7073 \\
    14  &  0.8115  &  0.2986  &  0.01230  &   2.70  &  2.287  &  0.8414  &  0.6828 \\
    16  &  0.7965  &  0.2779  &  0.01305  &   2.91  &  2.382  &  0.8309  &  0.6619 \\
    18  &  0.7832  &  0.2606  &  0.01361  &   3.10  &  2.470  &  0.8218  &  0.6437 \\
    20  &  0.7713  &  0.2460  &  0.01405  &   3.28  &  2.552  &  0.8138  &  0.6277 \\
    25  &  0.7457  &  0.2173  &  0.01478  &   3.69  &  2.736  &  0.7973  &  0.5946 \\
    30  &  0.7248  &  0.1961  &  0.01519  &   4.04  &  2.898  &  0.7842  &  0.5684 \\
    35  &  0.7071  &  0.1797  &  0.01544  &   4.36  &  3.044  &  0.7735  &  0.5469 \\
    40  &  0.6917  &  0.1665  &  0.01559  &   4.65  &  3.176  &  0.7644  &  0.5288 \\
    45  &  0.6782  &  0.1555  &  0.01568  &   4.91  &  3.299  &  0.7566  &  0.5131 \\
    50  &  0.6661  &  0.1463  &  0.01572  &   5.16  &  3.413  &  0.7497  &  0.4994 \\
    60  &  0.6453  &  0.1316  &  0.01575  &   5.61  &  3.620  &  0.7382  &  0.4763 \\
    70  &  0.6277  &  0.1202  &  0.01572  &   6.02  &  3.805  &  0.7287  &  0.4574 \\
    80  &  0.6125  &  0.1111  &  0.01567  &   6.38  &  3.974  &  0.7207  &  0.4414 \\
    90  &  0.5992  &  0.1036  &  0.01561  &   6.72  &  4.130  &  0.7139  &  0.4277 \\
   100  &  0.5873  &  0.0972  &  0.01555  &   7.04  &  4.275  &  0.7079  &  0.4157 \\
   150  &  0.5421  &  0.0761  &  0.01524  &   8.36  &  4.883  &  0.6859  &  0.3718 \\
   200  &  0.5106  &  0.0639  &  0.01499  &   9.42  &  5.368  &  0.6714  &  0.3428 \\
   250  &  0.4866  &  0.0557  &  0.01479  &  10.31  &  5.778  &  0.6608  &  0.3215 \\
   300  &  0.4673  &  0.0497  &  0.01464  &  11.10  &  6.136  &  0.6525  &  0.3049 \\
   350  &  0.4513  &  0.0451  &  0.01451  &  11.80  &  6.457  &  0.6457  &  0.2914 \\
   400  &  0.4376  &  0.0415  &  0.01440  &  12.44  &  6.749  &  0.6400  &  0.2801 \\
   450  &  0.4257  &  0.0385  &  0.01431  &  13.03  &  7.018  &  0.6352  &  0.2704 \\
   500  &  0.4152  &  0.0361  &  0.01423  &  13.58  &  7.267  &  0.6310  &  0.2620 \\
   600  &  0.3973  &  0.0321  &  0.01410  &  14.57  &  7.720  &  0.6240  &  0.2479 \\
   700  &  0.3826  &  0.0291  &  0.01400  &  15.47  &  8.125  &  0.6183  &  0.2365 \\
   800  &  0.3700  &  0.0267  &  0.01392  &  16.28  &  8.493  &  0.6135  &  0.2270 \\
   900  &  0.3592  &  0.0248  &  0.01385  &  17.02  &  8.832  &  0.6095  &  0.2189 \\
  1000  &  0.3496  &  0.0232  &  0.01379  &  17.72  &  9.146  &  0.6059  &  0.2119 \\
  2000  &  0.2911  &  0.0148  &  0.01347  &  22.96  &  11.52  &  0.5852  &  0.1703 \\
  3000  &  0.2603  &  0.0114  &  0.01333  &  26.65  &  13.18  &  0.5748  &  0.1496 \\
  4000  &  0.2400  &  0.0094  &  0.01325  &  29.59  &  14.51  &  0.5682  &  0.1364 \\
  5000  &  0.2252  &  0.0081  &  0.01320  &  32.07  &  15.62  &  0.5634  &  0.1269 \\
 10000  &  0.1838  &  0.0051  &  0.01307  &  41.10  &  19.68  &  0.5506  &  0.1012 \\
 20000  &  0.1491  &  0.0032  &  0.01299  &  52.49  &  24.80  &  0.5403  &  0.0806 \\
 30000  &  0.1317  &  0.0025  &  0.01296  &  60.48  &  28.39  &  0.5352  &  0.0705 \\
 40000  &  0.1205  &  0.0021  &  0.01294  &  66.85  &  31.26  &  0.5321  &  0.0641 \\
 50000  &  0.1124  &  0.0018  &  0.01293  &  72.22  &  33.69  &  0.5298  &  0.0596 \\
100000  &  0.0903  &  0.0011  &  0.01290  &  91.75  &  42.45  &  0.5236  &  0.0473 \\
\hline
		$N\to\infty$ & $4.387\,N^{-\frac13}$ & $2.406\,N^{-\frac23}$ & 0.01285~~ 
		& $2.040\,N^{\frac13}$ & \mbox{$0.9118\,N^{\frac13}$}
		& $\frac12{+}2K_{12}$ & $2.193\,N^{-\frac13}$ \\
		\hline
	\end{tabular}
\end{table}


\begin{thebibliography}{00}
\bibitem{note-omegaincr}
 The frequencies~\eqref{e.omegan} are in increasing order.
\bibitem{note-P0half}
 $\P_0$ has to be halved, but for simplicity of notation this is understood in the text: see the last paragraph of Appendix~\ref{a.density_rw}.
\bibitem{note-atan2}
 This expression of $\psi_k$ is useful for numerical calculations, e.g., using the function {\tt atan2(y,x)} whose range is in $[-\pi,\pi]$.
\end{thebibliography}
\end{document}